\documentclass[12pt]{article}
\usepackage{graphicx}
\usepackage{natbib}
\usepackage{amsmath,amssymb}
\usepackage{color}
\usepackage{textcomp}
\usepackage{subfigure}

\def\I{{\mathcal{I}}}
\def\J{{\mathcal{J}}}
\def\C{{\mathcal{C}}}

\def\X{{\mathcal{X}}}

\def\RR{{\mathbb{R}}}

\def\bA{{\bf A}}

\def\bS{{\bf S}}

\def\Em{E_{\textrm{max}}}

\def\a{{\mathbf a}}

\def\z{{\mathbf z}}
\def\f{{\mathbf f}}

\def\0{{\mathbf 0}}

\def\bomega{\boldsymbol{\omega}}
\def\bnabla{\boldsymbol{\nabla}}

\def\Dpartial#1#2{ {\partial #1 \over \partial #2} }

\def\Bmp#1{ \begin{minipage}{#1} }
\def\Emp{ \end{minipage} }
\def\Bmpc#1{ \begin{minipage}[c]{#1} }
\def\Bmpt#1{ \begin{minipage}[t]{#1} }
\def\Bmpb#1{ \begin{minipage}[b]{#1} }

\newtheorem{assumption}{Assumption}

\begin{document}
\title{Optimal Nonlinear Eddy Viscosity in Galerkin Models of Turbulent Flows} 

\author{Bartosz Protas$^1$\thanks{Email address for correspondence: bprotas@mcmaster.ca},
 Bernd R.~Noack$^2$ and 
 Jan \"Osth$^3$ \\ \\
$^1$Department of Mathematics and Statistics \\
McMaster University, Hamilton, ON, Canada
\\
$^2$Institut PPRIME, CNRS - Universit\'e de Poitiers - ENSMA, UPR 3346 \\
D\'epartment Fluides, Thermique, Combustion, CEAT \\ 
43 rue de l'A\'erodrome, F-86036 Poitiers CEDEX, France
\\
$^3$Division of Fluid Dynamics,
Department of Applied Mechanics \\
Chalmers University of Technology, SE-412 96 G\"oteborg, Sweden
}

\date{\today}

\maketitle

\begin{abstract}
  We propose a variational approach to identification of an optimal
  nonlinear eddy viscosity as a subscale turbulence representation for
  POD models.  The ansatz for the eddy viscosity is given in terms of
  an arbitrary function of the resolved fluctuation energy.  This
  function is found as a minimizer of a cost functional measuring the
  difference between the target data coming from a resolved direct or
  large-eddy simulation of the flow and its reconstruction based on
  the POD model.  The optimization is performed with a
  data-assimilation approach generalizing the 4D-VAR method.  POD
  models with optimal eddy viscosities are presented for a 2D
  incompressible mixing layer at $Re=500$ (based on the initial
  vorticity thickness and the velocity of the high-speed stream) and a
  3D Ahmed body wake at $Re=300,000$ (based on the body height and the
  free-stream velocity).  The variational optimization formulation
  elucidates a number of interesting physical insights concerning the
  eddy-viscosity ansatz used.  The 20-dimensional model of the
  mixing-layer reveals a negative eddy-viscosity regime at low
  fluctuation levels which improves the transient times towards the
  attractor.  The 100-dimensional wake model yields more accurate
  energy distributions as compared to the nonlinear modal
  eddy-viscosity benchmark {proposed recently} by \"Osth et al.\
  (2014).  Our methodology can be applied to construct quite arbitrary
  closure relations and, more generally, constitutive relations
  optimizing statistical properties of a broad class of reduced-order
  models.
\end{abstract}

\begin{flushleft}
\end{flushleft}



\section{Introduction}
\label{Sec:Introduction}

In this study we present an optimal nonlinear eddy-viscosity closure
for flow models based on the proper orthogonal decomposition (POD).
{We will focus on flows in unbounded domains which will be
  referred to here as ``open flows''.}  A reduced-order model (ROM)
may serve as a testbed for physical understanding of actual flow
phenomena, as a computationally inexpensive surrogate model for
optimization and as a low-order plant for control design.  The oldest
quantitative ROMs are vortex models which are over 100 years old
\citep[see, e.g.,][]{Lamb1945book}.  Most low-order vortex models of
open flows are hybrid systems with a heuristic account of the
creation, merging and annihilation of vorticity, and are thus not
amenable to most approaches of system reduction, stability analysis,
and control design.  Many current ROMs of fluid flows are based on the
traditional Galerkin method \citep[see, e.g.,][]{Fletcher1984book}.
In the kinematical step of this method, the flow {variables are}
expanded in terms of $N$ orthogonal basis functions
$\boldsymbol{u}_i$, $i=1,\dots,N$, as
$\boldsymbol{u}(\boldsymbol{x},t)= \sum_{i=1}^N a_i(t)
\boldsymbol{u}_i(\boldsymbol{x})$.  Thus, the mode coefficients $\a(t)
= [a_1(t), \dots, a_N(t)]^T \in \RR^N$ parameterize the fluid flow.
The dynamical step consists in representing the dependent variables in
the Navier-Stokes system in terms of such expansions and then
projecting on the individual modes which leads to the Galerkin system
in the general form
\begin{equation}
\frac{d\a}{dt} = \f(\a), \quad t > 0
\label{eq:ROM0}
\end{equation}
with propagator $\f \; : \; \RR^N \rightarrow \RR^N$.  Many ROMs
originate via post-processing of flow data obtained from simulations
or experiments and rely on the proper orthogonal decomposition
\citep[see, e.g.,][]{Noack2011book,Holmes2012book}.  In the following,
we focus on such POD models.

The error of the Galerkin model is expected to vanish for increasing
dimension $N$.  Since only a finite, and typically small, number of
modes is retained, this procedure results in a loss of information.
Hence, the reduced-order model \eqref{eq:ROM0} must be amended to
restore some physical features.  {Which features can be
  eliminated and which can be retained tends to depend on the nature
  of the particular problem.}  Generally, {however,} the
large-scale coherent structures with the associated production of
turbulent kinetic energy (TKE) are approximately resolved, while the
small-scale fluctuations responsible for most of the dissipation are
ignored.  The resulting excess production of the fluctuation energy
requires an additional stabilization in order to ensure the long-term
boundedness of solutions of system \eqref{eq:ROM0}.  The need to
introduce a suitable subscale turbulence representation gives rise to
a ``closure problem'' analogous to the problem encountered when
modeling turbulent flows based on the Reynolds-Averaged Navier-Stokes
(RANS) equations and Large-Eddy Simulations (LES), despite the fact
that the latter two approaches rely on flow descriptions in terms of
partial differential equations (PDEs), while system \eqref{eq:ROM0} is
finite-dimensional.  In particular, additional terms involving an
``eddy viscosity'' have been used in reduced-order models starting
with the pioneering work of \citet{Aubry1988jfm}.  These closure terms
have been refined in numerous studies leading to, e.g., the modal eddy
viscosities proposed by \citet{Rempfer1994jfm2}, calibration of an
auxiliary linear term investigated by \citet{Galletti2004jfm}, a
nonlinear term introduced by \citet{Cordier2013ef}, combinations
thereof studied by \citet{Osth2014jfm}, and projections of the
filtered Navier-Stokes equation \citep{Wang2011jcp}, just to name only
a few approaches.  In addition, projections onto more dissipative
subspaces were considered by \cite{Balajewicz2013jfm}.  We refer the
reader to \citet{wabi12} for some new proposals and a critical
assessment of several earlier approaches.

The discussed ROMs are all based on the Navier-Stokes equation.  In
principle, also the subscale closures can be {approximately modeled
  based on first-principle considerations} by means of structure and
parameter identification.  However, the availability of highly
resolved numerical and experimental data sets makes data-driven
modelling an appealing approach \citep[see,
e.g.,][]{Cacuci2013book,Kutz2013book}. {For example, in the context of
POD-based models, parameters of Galerkin systems and the required
closure relations can be accurately identified using variational
techniques of data assimilation \citep{Cordier2013ef}, collectively
known in the geosciences as ``4D-VAR'' \citep{k03}}.  A relatively
recent development is the construction of subscale turbulence models
based on optimization problems in which the closure model is adapted
using available measurements.  In the context of LES, this approach
has been pioneered by Moser et al.~leading to the concept of an
``optimal LES'' \citep{lm99}.  Optimization-based formulations of the
closure problem for Galerkin reduced-order models were recently
pursued in \cite{apma07,accm12,Cordier2013ef}.  In these studies the
authors obtained {\em time-dependent} eddy viscosities $\nu_T =
\nu_T(t)$ as minimizers of cost functionals representing the misfit
between the measured and reconstructed data.  However, the eddy
viscosity obtained in this way is a function of time and the
reduced-order model \eqref{eq:ROM0} is no longer autonomous.  {Since
  flow models with such time-dependent closures cannot be used to make
  predictions {\em outside} the time window on which the closure
  $\nu_T(t)$ was defined, this} limits the practical applicability of
such approaches.  In this context, we also mention the recent study by
\cite{hes14} in which an analogous time-dependent closure was obtained
for a vortex-based flow model.

In the present investigation we follow an optimization approach which
is fundamentally different: the optimal eddy viscosity is sought as a
function of the state $\a$, more precisely, its (turbulent)
fluctuation energy $E(t) := \| \a(t) \|^2_2=(1/2)\sum_{i=1}^N
a_i(t)^2$, so that the resulting ROM \eqref{eq:ROM0} will then be {\em
  autonomous}.  {Consequently, flow models with such closures
can be used to make predictions also outside the time window on which
the data assimilation was performed.}  The proposed reconstruction
approach is ``non-parametric'', in the sense that no assumptions are
made concerning the form of the dependence $\nu_T = \nu_T(E)$ other
than smoothness and the limiting behaviour for small and large values
of $E$. Relying on the concepts of data assimilation, the proposed
approach allows one to use measurement data in order to systematically
refine nonlinear eddy viscosity models obtained theoretically.
Therefore, it may be applicable to study the performance limitations
of a given ansatz for the eddy viscosity. The method builds on the
approach to the optimal reconstruction of constitutive relations in
complex multi-physics PDE problems developed by \citet{bvp10} and
\citet{bp11a}.  An application of this method to finite-dimensional
Galerkin models was carefully validated using a 3-state ROM of laminar
vortex shedding in the cylinder wake by \cite{pnm14}.  In the present
study, we employ this approach to identify optimal turbulence closures
in two medium and high-$Re$ flows, namely, a 2D incompressible mixing
layer and a 3D wake flow behind a blunt-back Ahmed body.  The
dimensions of the corresponding Galerkin models are $N=20$ for the
mixing layer and $N=100$ for the Ahmed body wake.  {As will be evident
  from the discussion below, these two flows exhibit distinct
  properties from the point of view of subgrid modelling and bear
  characteristics of, respectively, laminar and turbulent flows.}  In
addition to offering predictability improvements over existing
approaches \citep{Osth2014jfm}, the optimal turbulence closures also
reveal a number of unexpected yet physically plausible features, such
as negative values of the eddy viscosity in some ranges of the
turbulent kinetic energy $E$. We note that in fact the concept of a
negative eddy viscosity has already been invoked in the studies of
turbulent flows \citep[see, e.g.,][]{llgkt07a}.

The structure of the paper is as follows: In \S\ \ref{Sec:ROM} we
briefly recapitulate POD Galerkin models and highlight some properties
of the eddy viscosity in such models.  Our computational approach is
outlined in \S\ \ref{Sec:OptimalEddyViscosity}.  Optimal eddy
viscosities and the properties of the resulting ROMs of the mixing
layer and the Ahmed body flow are presented and analyzed in \S\
\ref{Sec:Results}.  Summary and future directions are provided in \S\
\ref{Sec:Conclusions}, whereas some technical material concerning the
optimization approach is collected in Appendix \ref{Sec:Adjoint}.

\section{POD modeling}
\label{Sec:ROM}
In this section, POD models for turbulent flows are briefly reviewed.
First (\S\ \ref{Sec:Configuration}), the assumed flow configurations
are specified.  The POD expansion and the corresponding Galerkin
projection of the Navier-Stokes equation are described in \S\
\ref{Sec:POD} and \S\ \ref{Sec:GalerkinProjection}, respectively.  In
\S\ \ref{Sec:EddyViscosity}, a nonlinear eddy viscosity ansatz is
introduced against which the optimal relations of the next section
will be benchmarked.  Finally (\S\ \ref{Sec:TransientTimes}),
conditions for the appearance of negative values of eddy viscosity are
identified thus setting the stage for the optimization formulation of
\S\ \ref{Sec:OptimalEddyViscosity} and the initially somewhat
surprising results reported in \S\ \ref{Sec:Results}.

\subsection{Flow configurations}
\label{Sec:Configuration}
We assume an incompressible flow of a Newtonian fluid in a stationary
domain $\Omega$.  The fluid is described by the density $\rho$ and
kinematic viscosity {$\tilde{\nu}$}.  The position and time are
denoted $\boldsymbol{x}$ and $t$, respectively.  The flow field is
described by the velocity $\boldsymbol{u}$ and pressure $p$.  The
fluid motion is characterized by a velocity scale $U$ and a length
scale $L$, {which will take different numerical values in the problems
studied here,} and define the Reynolds number {as $Re:=U L /
  {\tilde{\nu}}$}.  In the following, all quantities are assumed to be
non-dimensionalized by $U$, $L$ and $\rho$, and $\nu:=1/Re$ represents
the reciprocal Reynolds number {(``$:=$'' means that the left-hand
  side of the equation is defined by the right-hand side)}.  The fluid
motion is governed by the continuity equation and the momentum balance

\begin{subequations}
\label{Eqn:NavierStokes}
\begin{align}
\label{Eqn:Continuity}
\nabla \cdot \boldsymbol{u} & = 0, \\
{\Dpartial{\boldsymbol{u}}{t}} + \boldsymbol{u} \cdot \nabla \boldsymbol{u}  & = - \nabla p + \nu \triangle \boldsymbol{u} 
\end{align}
\end{subequations}
subject to suitable initial and boundary conditions.

\begin{figure}
\begin{center}
\includegraphics[width=0.8\textwidth]{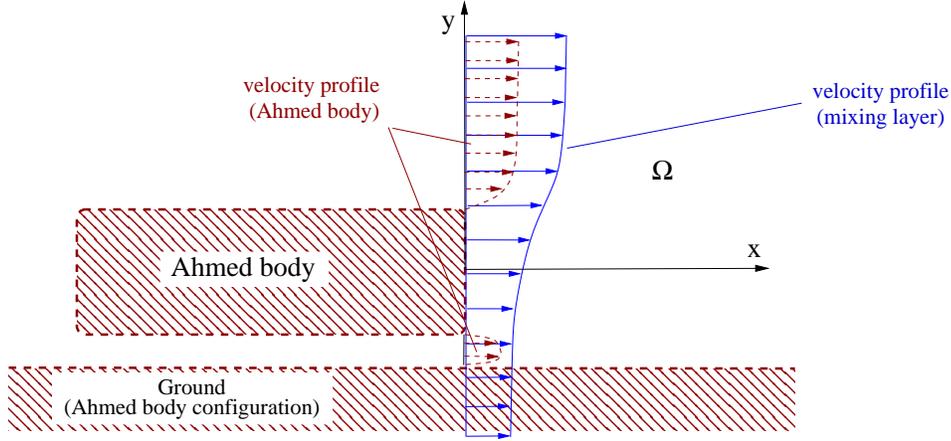}
\caption{{Schematic of the coordinate system used in the example
    problems studied here. The velocity profile in the mixing-layer
    problem is indicated with blue solid lines, whereas the Ahmed body
    configuration together with the corresponding mean velocity
    profiles are marked with dark red dashed lines.}}
\label{fig:domain}
\end{center}
\end{figure}

While the proposed methodology is fairly general, to fix attention, in
this study we investigate two shear flows, a 2D spatially evolving
mixing layer with a narrow frequency bandwidth and a 3D wake behind an
Ahmed body with a broad frequency bandwidth including a slow drift of
the base flow.  In both flows, the origin of the Cartesian coordinate
system is at the center of the inlet of the observation domain, i.e.,
is located at the maximum shear position in case of the mixing layer
and at the center of the rear face of the Ahmed body {(figure
\ref{fig:domain})}.  The $x$-axis points in the direction of the flow,
the $y$-axis is aligned with the shear and the $z$-axis is orthogonal
to the $x$- and $y$-coordinates.

\subsection{Proper orthogonal decomposition}
\label{Sec:POD}

We perform a POD expansion \citep{Lumley1970book} of $M$ velocity
snapshots $\boldsymbol{u}^m := \boldsymbol{u}(\boldsymbol{x},t^m)$
sampled at equispaced time instances $t^m = m \Delta t$,
$m=1,\ldots,M$, with the time step $\Delta t$.  The averaging
operation of any velocity-dependent function $\boldsymbol{F} (
\boldsymbol{u} )$ over this ensemble is denoted by an overbar, i.e.,
\begin{equation}
\label{Eqn:Average}
\overline{ \boldsymbol{F} ( \boldsymbol{u} )} 
:= \frac{1}{M} \sum_{m=1}^M \boldsymbol{F} \left( \boldsymbol{u}^m \right).
\end{equation}
The inner product for two velocity fields
{${\z_1, \z_2} \in L_2 ( \Omega )$} is defined as
\begin{equation}
\label{Eqn:InnerProduct}
{\big\langle \z_1, \z_2 \big\rangle_{L_2(\Omega)} :=
\int\limits_{\Omega} \, \z_1 \cdot \z_2 \, d\boldsymbol{x}.}
\end{equation}
This inner product defines the energy norm
$\Vert \boldsymbol{u} \Vert_{L_2(\Omega)} = \sqrt{\langle \boldsymbol{u}, \boldsymbol{u} \rangle_{L_2(\Omega)}}$.

The averaging operation and the inner product uniquely define the
corresponding snapshot POD \citep{Sirovich1987qam1,Holmes2012book}.
First, following the Reynolds decomposition, the velocity field is
decomposed into a mean field $\boldsymbol{u}_0$ and a fluctuating
contribution $\boldsymbol{u}'$ defined as
\begin{equation}
\boldsymbol{u}_0 := \overline{\boldsymbol{u}}, \qquad \boldsymbol{u}' := \boldsymbol{u} - \overline{\boldsymbol{u}}.
\label{Eqn:u0}
\end{equation}
Then, the fluctuating part is approximated by a Galerkin expansion
with space-dependent modes $\boldsymbol{u}_i(\boldsymbol{x})$,
$i=1,2,\ldots,N$, used as the basis functions and the corresponding
mode coefficients $a_i(t)$
\begin{equation}
 \label{Eqn:POD+Residual}
 \boldsymbol{u(\boldsymbol{x}},t) = \boldsymbol{u}_0(\boldsymbol{x})
 +  \sum_{i=1}^{N} a_i(t) \boldsymbol{u}_i(\boldsymbol{x}) + \boldsymbol{u}_{res}(\boldsymbol{x},t),
\end{equation}
where $\boldsymbol{u}_{res}$ represents the residual.  POD yields a
Galerkin expansion with the minimal average squared residual
$\overline{\left\Vert \boldsymbol{u}_{res}
  \right\Vert^2_{L_2(\Omega)}}$ as compared to any other Galerkin
expansion with $N$ modes \citep{Lumley1970book}.  We note that the
snapshot POD method limits the number of POD modes to $N \le M-1$.

To facilitate subsequent developments, we rewrite the POD expansion
more compactly following the convention of
\citet{Rempfer1994jfm,Rempfer1994jfm2}:
\begin{equation}
 \label{Eqn:POD}
 \boldsymbol{u}(\boldsymbol{x},t) =  \sum_{i=0}^{N} a_i(t) \boldsymbol{u}_i(\boldsymbol{x}),
\end{equation}
where $a_0(t) \equiv 1$ (because of this property we will refer to the phase
space as $N$-dimensional, even though the state vector $\a(t)$ has
formally the dimension $N+1$). For later reference, we recapitulate
the first and second moments of the POD mode coefficients:
\begin{equation}
\label{Eqn:POD_Statistics}
\overline{a_i} = 0, \quad \overline{a_i a_j} = \lambda_i \delta_{ij}, \quad
i,j = 1,\dots,N,
\end{equation}
where $\lambda_i$ are the POD eigenvalues. The energy content in each
mode is given by $E_i(t) := a_i^2(t)/2$ and the turbulent kinetic
energy resolved by the Galerkin expansion $E(t)$ is
\begin{equation}
 \label{Eqn:TotalK}
 E(t) =\sum_{i=1}^{N} E_i(t).
\end{equation}
At any fixed time $t$, the limit $\lim_{N \to \infty} E(t)$ for POD yields the total turbulent
kinetic energy $K(t)$ of the original velocity field.
We note that, by \eqref{Eqn:POD_Statistics}, the average modal energy and POD
eigenvalues are synonymous: $\overline{E_i} = \lambda_i/2$.

\subsection{Galerkin projection}
\label{Sec:GalerkinProjection}

The Galerkin expansion \eqref{Eqn:POD} satisfies the incompressibility
condition and the boundary conditions by construction.  The evolution
equation for the mode coefficients $a_i$ is derived by a
\textit{Galerkin projection} of the Navier-Stokes equation
\eqref{Eqn:NavierStokes}, written in the operator form as
$\boldsymbol{R} (\boldsymbol{u}) = \boldsymbol{0}$, onto individual
POD modes, i.e., via $\left \langle \boldsymbol{u}_i, \boldsymbol{R}
  (\boldsymbol{u}) \right\rangle_{L^2(\Omega)} = 0$, $i=1,\dots,N$.
Details are provided in the monographs by \citet{Noack2011book} and
\citet{Holmes2012book}.  For internal flows, the Galerkin
representation of the pressure term vanishes.  For open flows with
large domains and three-dimensional fluctuations, the pressure term
can generally be neglected as discussed by \citet{Deane1991pfa},
\citet{Ma2002jfm} and \citet{Noack2005jfm}.  Here, the Galerkin
projection of the pressure term was found to be negligible and it is
therefore omitted from the model.  Thus, the Galerkin system
describing the temporal evolution of the modal coefficients, $a_i(t)$,
reads
\begin{equation}
 \label{Eqn:GalerkinSystem}
 \frac{da_i}{dt} = f_i(\boldsymbol{a}) = \nu \sum_{j=0}^{N} l_{ij}^{\nu} a_j + \sum_{j,k = 0}^{N} q_{ijk}^c a_j a_k, \quad i=1,\dots,N.
\end{equation}
The coefficients $l_{ij}^{\nu}$ and $q_{ijk}^c$, $i,j,k=0,\ldots,N$,
are the Galerkin coefficients describing, respectively, the viscous
and convective effects in the Navier-Stokes system
\eqref{Eqn:NavierStokes}.  For internal flows with the Dirichlet or
periodic boundary conditions, the quadratic term can be shown to be
exactly energy-preserving
\begin{equation}
\label{Eqn:EnergyPreservationQijk}
q_{ijk}^c + q_{ikj}^c + q_{kij}^c + q_{kji}^c + q_{ikj}^c + q_{jik}^c = 0, \quad
i,j,k=1,\ldots,N. 
\end{equation}
{Energy preservation \eqref{Eqn:EnergyPreservationQijk} can be also be
proven for flows past obstacles in unbounded domains under the
condition that the velocity fluctuations decay at infinity. For finite
domains, relation \eqref{Eqn:EnergyPreservationQijk} is still a good
approximation assuming that the fluctuations have significantly
decreased at the downstream boundary, as is the case for the cylinder
wake example discussed below. Even when more significant fluctuation
levels are present at the downstream boundary as in the mixing layer
flow, the enforced anti-symmetry of $q_{ijk}$ is numerically found not
to noticeably change the behaviour of the POD model in the examples
considered.}

\subsection{Post-transient fluctuation levels}
\label{Sec:EddyViscosity}

For turbulent flows, POD models face one well-known challenge
addressed already in the pioneering work of \citet{Aubry1988jfm}: the
finite POD expansion often contains a fraction of the total
fluctuation energy.  While a significant portion of the TKE production
may be resolved by the large-scale structures contained in the POD
expansion, most of the dissipation in the small-scale eddies is
ignored in the Galerkin system.  The resulting over-production of TKE
in the POD model leads to over-prediction of the fluctuation level,
including possible divergence to infinity in finite time.  A common
cure is the inclusion of an ``eddy viscosity'' term absorbing the
excess energy,
\begin{equation}
\label{Eqn:GalerkinSystemEddyViscosity}
 \frac{da_i}{dt} = f_i(\boldsymbol{a}) + \nu_T  \sum_{j=0}^N l_{ij}^{\nu} a_j,
\quad i=1,\dots,N.
\end{equation}
Generally, off-diagonal elements $l_{ij}^{\nu}$, $i \not = j$, are
small and therefore negligible.

In early studies eddy viscosity $\nu_T$ was assumed to be a constant
parameter.  Yet, the non-physical implication is that the POD-resolved
part of the turbulent flow effectively behaves like a laminar flow
with reciprocal Reynolds number $\nu_{\tiny\rm eff} = \nu + \nu_T$.
Another non-physical implication is that a {\em linear} Galerkin term
is to represent the {\em nonlinear} energy cascade.  Numerous
refinements of this eddy viscosity term have been suggested as
discussed by \citet{Osth2014jfm}.  {To simplify the notation, hereafter
we will use the convention that the superscript symbol ``$^\circ$''
will denote quantities related to closure models {obtained based
  on} theoretical arguments, whereas the superscript symbol
``$^\bullet$'' will denote the corresponding quantities related to
closure models derived from actual data.}  In this study, our point of
departure is a nonlinear modal eddy viscosity
\begin{equation}
\label{Eqn:ReferenceEddyViscosity}
 \nu_T^{\circ} {:=} \nu_T^{a}  \> \sqrt{\frac{E(t)}{\overline{E}}} \, \kappa_i
\end{equation}
with the mode-dependent factor $\kappa_i$, $i=1,\dots,N$.  This factor
is equal to the unity, $\kappa_i \equiv 1$, for the global
eddy-viscosity ansatz and is derived from the modal power balance of
the flow \citep{Noack2005jfm} for the modal eddy viscosity.  {The
quantity $\nu_T^{a}$ represents {a constant} reference value of
the eddy viscosity obtained from a long-time average of energy
dissipation in the flow on the attractor, where the latter is defined
as usual in dynamical systems as a subset of the phase space to which
all trajectories converge regardless of the initial positions. Thus,
the eddy viscosity $\nu_T^{\circ}$ {defined in
  \eqref{Eqn:ReferenceEddyViscosity}} becomes larger than the
reference value $\nu_T^{a}$ when the instantaneous resolved
fluctuation energy $E(t)$ exceeds $\overline{E}$ and vice versa}.  The
square-root dependency of $\nu_T^{\circ}$ on $E(t)$ is motivated by a
scaling argument \citep{Noack2011book} and we add that this nonlinear
eddy viscosity term guarantees the boundedness of any Galerkin
solution \citep{Cordier2013ef}. {Hereafter we will refer to
  relation \eqref{Eqn:ReferenceEddyViscosity} as the ``reference eddy
  viscosity''.}

\subsection{Transient dynamics}
\label{Sec:TransientTimes}

The nonlinear eddy viscosity term effectively has the ability to
prevent non-physically large fluctuation levels.  Another frequently
observed shortcoming of POD systems are significantly over-predicted
transient times, even for laminar flows.  To shed light on this issue
and show how it can be remedied through a suitable choice of a
nonlinear eddy viscosity, in the following we consider one of the
simplest POD Galerkin models exhibiting non-physical transient times
\textit{and} non-physical fluctuation levels.  The starting point is
the 2D laminar cylinder wake at $Re=100$ in an unbounded domain
truncated for computational purposes to a finite box
\citep{Noack2003jfm}.  The first two POD modes resolve already 95\% of
the fluctuation energy and we chose $N=2$ as the model order.  The POD
system is effectively phase-invariant and is well approximated by a
linear oscillator:
\begin{subequations}
\label{Eqn:WakeModel}
\begin{eqnarray}
{\frac{da_1}{dt}} &=& f_1(a_1,a_2) = \sigma^{\circ} a_1 - \omega^{\circ} a_2, \\
{\frac{da_2}{dt}} &=& f_2(a_1,a_2) = \sigma^{\circ} a_2 + \omega^{\circ} a_1, \\
\sigma^{\circ} &=& 0.0073, \label{Eqn:WakeModelc} \\
\omega^{\circ} &=& 1.0763,
\end{eqnarray}
\end{subequations}
{which is obtained through a standard Galerkin projection
  procedure (see \cite{Noack2003jfm} for details and validation).}
The quadratic term vanishes by \eqref{Eqn:EnergyPreservationQijk} and
the observed phase invariance.  Evidently, \eqref{Eqn:WakeModel}
describes an oscillatory behaviour with a slow exponential growth,
i.e., growth without bound.

The mode coefficients $a_i^{\bullet}$, $i=1,2$, obtained from a direct
numerical simulation (DNS) starting from the steady solution quickly
converge to a limit cycle.  This transient is far better approximated
by the following mean-field model exhibiting a stable limit cycle at
$r_{\infty} \approx 2.3$ \citep{pnm14}:
\begin{subequations}
\label{Eqn:MeanFieldModel}
\begin{eqnarray}
{\frac{da_1}{dt}} &=& \sigma^{\bullet} a_1 - \omega^{\bullet} a_2, \\ 
{\frac{da_2}{dt}} &=& \sigma^{\bullet} a_2 + \omega^{\bullet} a_1, \\
\sigma^{\bullet} &=& \sigma_1 \left[1 -  r^2 / r_{\infty}^2 \right], \\
\omega^{\bullet} &=& \omega_1   + 0.150 \> r^2 / r_{\infty}^2.
\end{eqnarray}
\end{subequations}
with $r := \sqrt{a_1^2+a_2^2}$, and $\sigma_1 = 0.151$ and
$\omega_1=0.886$ representing the initial {(i.e., evaluated at
  the unstable fixed point)} growth rate and frequency of the
transient solution {(these values are obtained via calibration against
the DNS data)}.

\begin{figure}
\begin{center}
\mbox{
\subfigure[]{\includegraphics[width=0.5\textwidth]{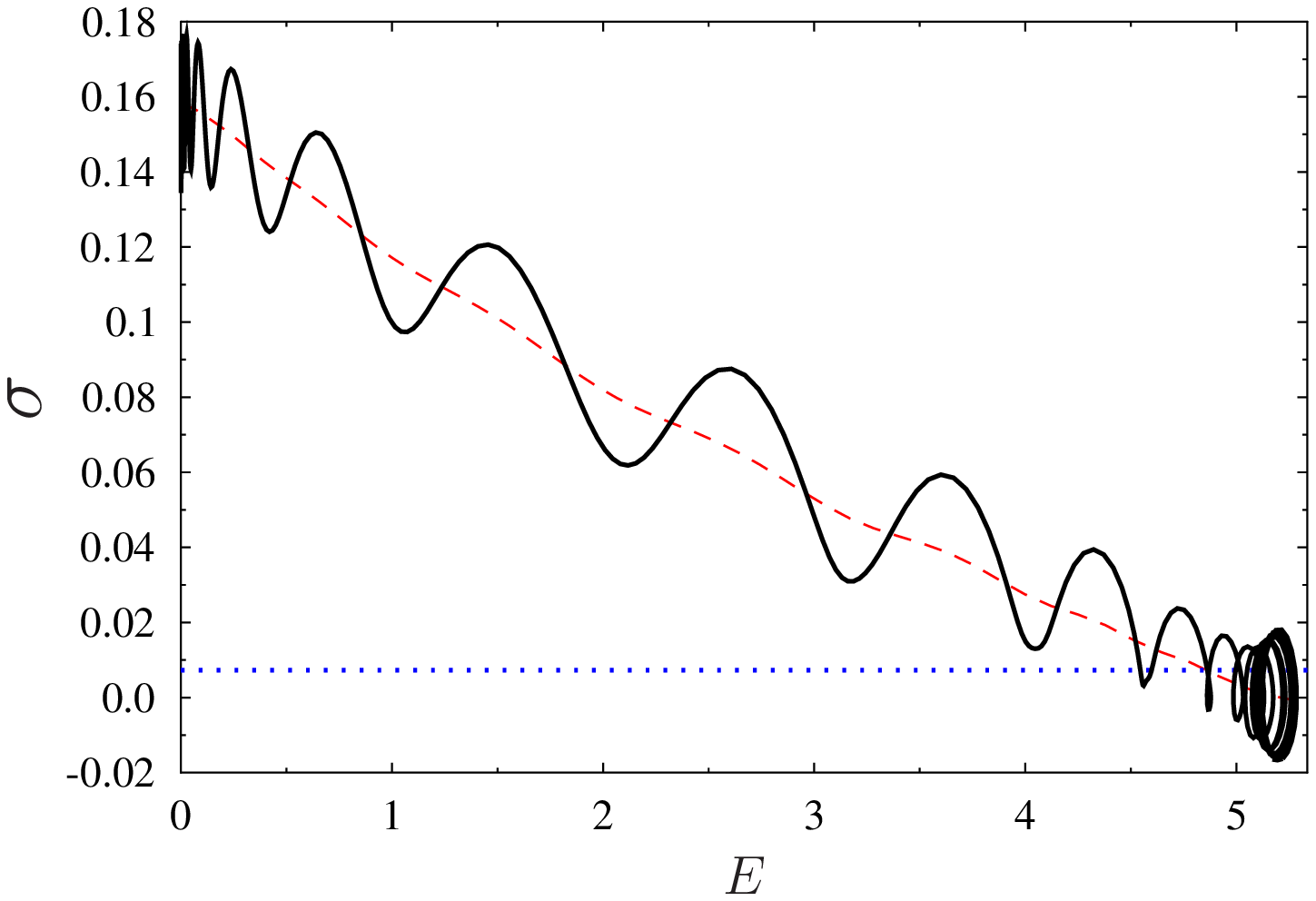}}
\subfigure[]{\includegraphics[width=0.5\textwidth]{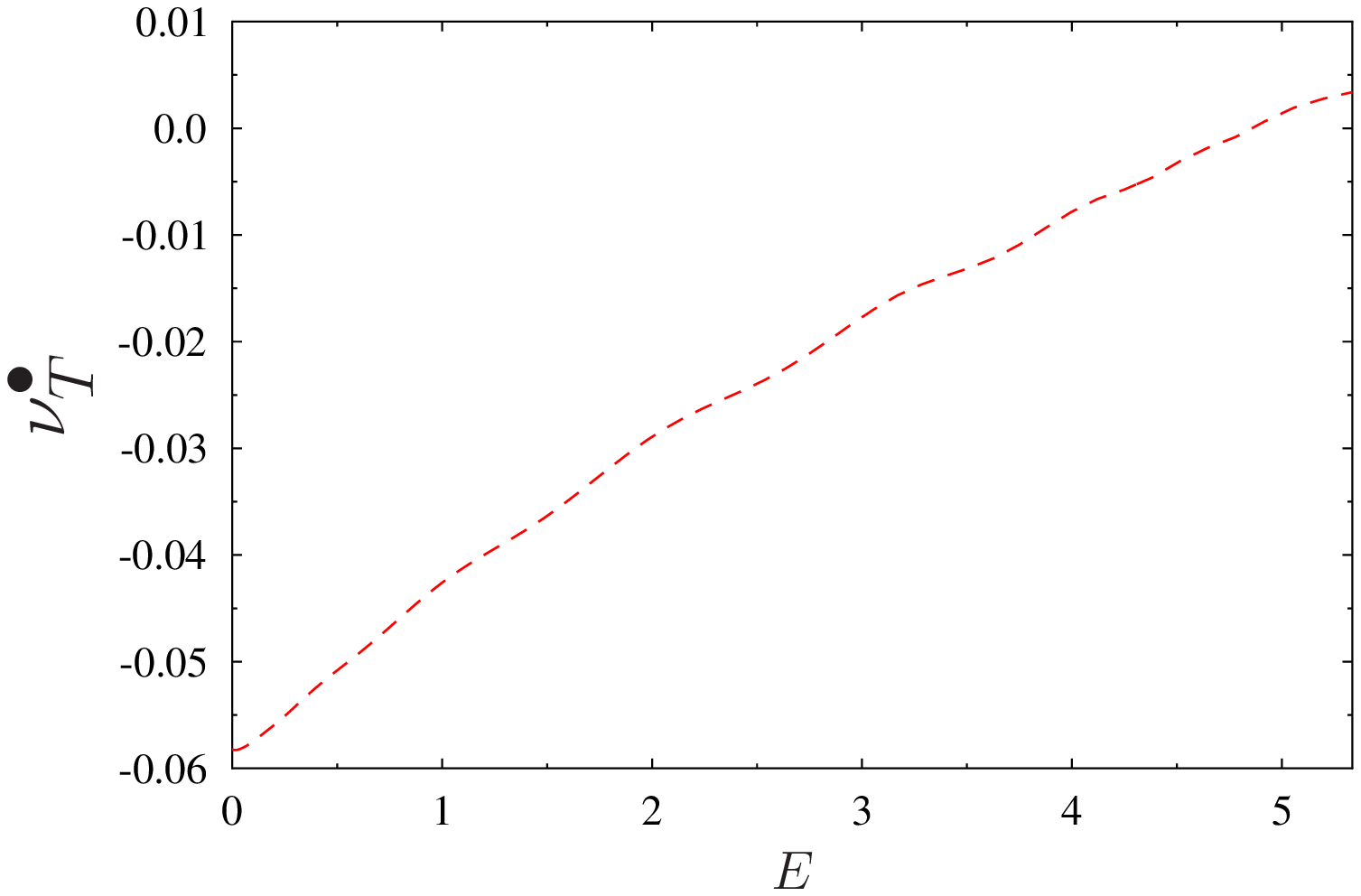}}}
\caption{Results obtained for the two-dimensional cylinder wake flow
  at $Re=100$: (a) the growth rates $\sigma$ in reduced-order models
  \eqref{Eqn:WakeModel}--\eqref{Eqn:MeanFieldModel} and (b) the
  corresponding optimal eddy viscosity $\nu_T^{\bullet}$ in system
  \eqref{Eqn:WakeModelViscosity} as functions of the fluctuation
  energy $E$; optimal reconstructions $\sigma^{\bullet}$ and
  $\nu_T^{\bullet}$ {computed by \cite{pnm14}} (red dashed
  lines), prediction from Galerkin model \eqref{Eqn:WakeModel} (blue
  dotted line) and the quantity $r^{-1} \, (dr/dt)|_{r(t)}$ computed
  based on the solution of the Navier-Stokes problem (black solid
  line).}
\label{fig:g1l}
\end{center}
\end{figure}

The growth rate \eqref{Eqn:WakeModelc} of the POD model is thus
initially underpredicted by more than a factor of $20$ while it is
increasingly overpredicted near and beyond the limit cycle.  We aim to
correct this growth rate using the eddy viscosity ansatz of the form
\eqref{Eqn:GalerkinSystemEddyViscosity} which results in:
\begin{subequations}
\label{Eqn:WakeModelViscosity}
\begin{eqnarray}
{\frac{da_1}{dt}} &=& f_1(a_1,a_2) + \nu_T l_{11}^{\nu} a_1, \\
{\frac{da_2}{dt}} &=& f_2(a_1,a_2) + \nu_T l_{22}^{\nu} a_2.
\end{eqnarray}
\end{subequations}
Here, $ l_{11}^{\nu} = l_{22}^{\nu} <0$ by the assumed phase
invariance and the dissipativity property of the viscous term.
Matching the growth rate of \eqref{Eqn:WakeModelViscosity} with the
DNS-inferred mean-field model \eqref{Eqn:MeanFieldModel} yields
\begin{displaymath}
{\sigma^{\circ} + \nu_T \, \l_{11}^{\nu}  = \sigma^{\bullet}
\quad \Longrightarrow \quad 
\nu_T  \,  l_{11}^{\nu} = \sigma^{\bullet}- \sigma^{\circ} 
=\sigma_1 \left[ 1 - r^2/r_{\infty}^2  \right]  -\sigma^{\circ}.}
\end{displaymath}
Evidently, the eddy viscosity is an affine function of the fluctuation
energy $E(t)= r(t)^2/2$, i.e., 
\begin{equation}
\label{Eqn:WakeEddyViscosity}
\nu_T(E) = a + b E
\end{equation}
with a negative intercept $a = ({\sigma_1 - \sigma^{\circ}}) /
l_{11}^{\nu}$ and a positive slope $b = {-}\sigma_1/(E_\infty \,
l_{11}^{\nu})$, in which $E_\infty = \overline{(a_1^2 + a_2^2)}/2 =
r_\infty^2/2$, so that $\nu_T = - \sigma_1/l_{11}^{\nu} > 0$ at $E =
E^{a}$, where $E^a$ is the fluctuating energy level corresponding to
the attractor.  {Different aspects of these observations are
  illustrated in figure \ref{fig:g1l}.  In addition to the growth rate
  predicted by the standard POD model \eqref{Eqn:WakeModelc} and the
  growth rate $r^{-1} \, (dr/dt)|_{r(t)}$ characterizing the DNS of
  the actual Navier-Stokes flow, in figure \ref{fig:g1l}a we also show
  the optimal growth rate $\sigma^{\bullet}(E)$ reconstructed by
  \cite{pnm14} using a similar methodology as employed in the present
  study. It is clear from this figure that the optimal growth rate
  depending on the fluctuating energy provides a much better
  representation of the actual data than does the constant growth rate
  produced by the Galerkin procedure.  The eddy viscosity
  $\nu_T^{\bullet}$ corresponding to the optimal growth rate
  $\sigma^{\bullet}$ is shown as a function of $E$ in figure
  \ref{fig:g1l}b {(this data is not shown for system
    \eqref{Eqn:MeanFieldModel}, because it does not explicitly involve
    a term with eddy viscosity, hence $\nu_T^{\bullet}$ is not defined
    in that case)}.  The key message from this figure is that the form
  of the optimally reconstructed eddy viscosity is quite similar to
  \eqref{Eqn:WakeEddyViscosity} and features both positive and
  negative values.  We also remark here} that the form of
\eqref{Eqn:WakeEddyViscosity} as an affine function of $E$ is
different from \eqref{Eqn:ReferenceEddyViscosity} which involves a
square-root expression.  There is, however, no contradiction, since
\eqref{Eqn:ReferenceEddyViscosity} is obtained for the flow energy
cascade with triadic mode interactions, while the mean-field model
\eqref{Eqn:MeanFieldModel} describes the change of the growth rate due
to base-flow variations with the associated Reynolds stresses
proportional to $E$.

Summarizing, a negative eddy viscosity at low fluctuation values
{and positive at large fluctuation values} can cure
non-physically long transient times to the attractor.  In the
following we thus allow the eddy viscosity to be an essentially
arbitrary function of $E$
\begin{equation}
\nu_T^{\bullet} := \nu_T^{\bullet} \left(E \right).
\end{equation}
In the light of the cylinder wake example, one may therefore expect
small or negative values of the eddy viscosity to arise for $E <
\overline{E}$ and positive values for $E > E^{a}$.

The marginal growth rates of POD models may be also related to
unresolved base flow variations
\citep{Aubry1988jfm,Podvin2009pf,Noack2003jfm} and mode deformation
during transients \citep{Noack2003jfm,Sapsis2013pnas}. While it is
possible to address these issues in our framework, it would
significantly complicate the exposition, hence they will not be
considered in the present study.

\section{Optimal eddy viscosity}
\label{Sec:OptimalEddyViscosity}

\begin{figure}
\begin{center}
\includegraphics[width=0.8\textwidth]{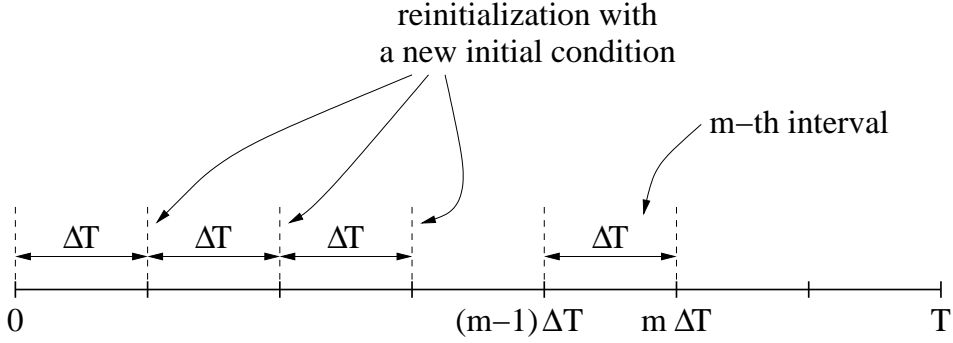}
\caption{Schematic showing the partition of the time window $[0,T]$
  into subintervals $[(m-1)\Delta T, m\Delta T]$, $m=1,\dots,M$.}
\label{fig:subintervals}
\end{center}
\end{figure}

In this section we describe a variational approach for determination
of an {\em optimal} dependence of the nonlinear eddy viscosity $\nu_T$
in the Galerkin system \eqref{Eqn:GalerkinSystemEddyViscosity} on the
turbulent kinetic energy $E$. Here, ``optimality'' means that the eddy
viscosity minimizes a performance criterion quantifying how well the
evolution described by reduced-order model
\eqref{Eqn:GalerkinSystemEddyViscosity} matches the actual evolution
governed by Navier-Stokes system \eqref{Eqn:NavierStokes}. {We
  consider a time window $[0,T]$ whose length $T$ is a parameter and}
assume that over {this} time window the flow is characterized by
the resolved turbulent kinetic energy $\widetilde{E}(t)$ representing
the energy content of its first $N$ POD modes, i.e.,
\begin{equation}
\widetilde{E}(t) := \frac{1}{2}\,
\sum_{i=1}^N \langle\boldsymbol{u}'(\cdot,t),\boldsymbol{u}_i\rangle_{L_2(\Omega)}^2,
\label{Eqn:tKsig}
\end{equation}
{where $\boldsymbol{u}'$, $\boldsymbol{u}_i$ and the inner
  product $\langle\cdot,\cdot\rangle_{L_2(\Omega)}$ were defined in \S\
  \ref{Sec:POD}.}  This fluctuation energy is determined from the
solution (here, DNS or LES) of the Navier-Stokes problem.  Then, we
can define the following cost functional
\begin{equation}
  \J(\nu_T) =  \frac{1}{2T}\int_0^T \left[ E(t;\nu_T) - 
\widetilde{E}(t) \right]^2\, dt,
\label{Eqn:J}
\end{equation}
where $E(t;\nu_T)$ is the turbulent kinetic energy characterizing
system \eqref{Eqn:GalerkinSystemEddyViscosity} which depends on eddy
viscosity $\nu_T$. Since the length $T$ of the time window on which
measurements $\widetilde{E}(t)$ are available can be quite long
compared to the times over which the reduced-order model
\eqref{Eqn:GalerkinSystemEddyViscosity} is capable of reproducing
accurately the actual trajectory, in evaluating $E(t;\nu_T)$ we will
periodically restart system \eqref{Eqn:GalerkinSystemEddyViscosity}
using projections of the actual flow evolution on the POD modes as the
initial data $\a^0$.  More precisely, we will subdivide the interval
$[0,T]$ into $M$ subintervals of length $\Delta T = T / M$, so that
$[0,T] = [0,\Delta T]\cup[\Delta T, 2 \Delta
T]\cup\ldots\cup[(M-1)\Delta T, M\Delta T]$, see figure
\ref{fig:subintervals}. On each of the subintervals $[(m-1)\Delta T,
m\Delta T]$, $m=1,\dots,M$, the Galerkin system will therefore take
the form
\begin{subequations}
\label{Eqn:GalerkinSystem4}
\begin{align}
& \frac{da_i}{dt} =  \sum_{j,k=0}^N q^c_{ijk} a_j a_k 
+ \left[ \nu + \nu_T(E(t))\right] \, \sum_{j=0}^N l^{\nu}_{ij} a_j, \quad t \in ((m-1)\Delta T, m\Delta T], \label{Eqn:GalerkinSystem4a} \\
& a_i((m-1)\Delta T) = a^{0,m}_i, \qquad i=1,\dots,N, 
\label{Eqn:GalerkinSystem4b}
 \end{align}
\end{subequations}
where $a^{0,m}_i = \langle\boldsymbol{u}'(\cdot,(m-1)\Delta
T),\boldsymbol{u}_i\rangle_{L_2(\Omega)}$ and $m=1,\dots,M$.
{Periodic restarts of Galerkin system \eqref{Eqn:GalerkinSystem4}
  ensure that its trajectory never {departs} too far from the
  projected trajectory of the actual flow, which is important given
  the form of the cost functional adopted in \eqref{Eqn:J}.}

The nonlinear eddy viscosity $\nu_T^{\circ}$ introduced in \S\
\ref{Sec:EddyViscosity}, cf.~\eqref{Eqn:ReferenceEddyViscosity}, will
serve as a reference and point of departure for the present
optimization approach. As regards the functional form of the optimal
eddy viscosity $\nu_T^{\bullet}$, we will make the following rather
nonrestrictive assumptions (hereafter we will use the symbol $e$ as the
variable corresponding to the turbulent kinetic energy $E$).
\begin{assumption}
\begin{enumerate}
\item $\nu_T^{\bullet}(e)$ is defined for $e \in \I := [0,\Em]$, where
  $\Em$ is chosen such that $\Em > \max_{t \in [0,T]} E(t)$,

\item $\nu_T^{\bullet}(e)$ is a continuous function of $e$ with
  square-integrable derivatives on $\I$; this implies that
  $\nu_T^{\bullet} \in H^1(\I)$, where $H^1(\I)$ is the Sobolev
  function space equipped with the inner product \citep{af05}
\begin{equation}
\forall_{z_1,z_2 \in H^1(\I)} \qquad
\big\langle z_1, z_2 \big\rangle_{{H^1(\I)}} = 
\int_0^{\Em} z_1 z_2 + \ell^2 \Dpartial{z_1}{e} \Dpartial{z_2}{e}\, de,
\label{Eqn:ipH1}
\end{equation}
where $\ell > 0$,

\item
\begin{equation}
\nu_T^{\bullet}(0) = \nu_T^{\circ}(0) = 0,
\label{Eqn:nuTK0}
\end{equation}

\item
\begin{equation}
\frac{d \nu_T^{\bullet}}{de}\bigg|_{e=\Em} = \frac{d \nu_T^{\circ}}{de}\bigg|_{e=\Em} =: G.
\label{Eqn:DnuTKm}
\end{equation}
\end{enumerate}
\label{ass1}
\end{assumption}
\noindent 
{Some comments are in place as regards the physical
  interpretation of the above assumptions. Assumption \ref{ass1}(a)
  guarantees that the optimal eddy viscosity $\nu_T^{\bullet}(e)$ is
  defined over a range of $e$ relevant for the given flow. {Our
    experience shows that the specific value of $\Em$ does not
    noticeably influence the results, provided it is slightly larger
    than $\max_{t \in [0,T]} E(t)$, typically by a factor in the range
    1.1--3.0.}  Assumption \ref{ass1}(b) concerns the minimal
  smoothness of the optimal eddy viscosity as a function of $e$.  We
  emphasize that, as shown by \cite{bvp10}, omitting the
  differentiability requirement and assuming that $\nu_T^{\bullet}$ is
  only square-integrable ($\nu_T^{\bullet} \in L_2(\I)$) could in fact
  produce {\em discontinuous} eddy viscosities which are unphysical.
  Assumptions \ref{ass1}(c) and \ref{ass1}(d) imply that for limiting
  values of $e$ the behaviour of the optimal eddy viscosity
  $\nu_T^{\bullet}$ is the same as in the reference relation
  \eqref{Eqn:ReferenceEddyViscosity}. More specifically, at $e=0$ the
  optimal eddy viscosity $\nu_T^{\bullet}$ will vanish, whereas at $e
  = \Em$ it will have the same slope (with respect to $e$) as the
  reference relation $\nu_T^{\circ}$. The latter assumption is
  justified by the fact, shown by {\cite{Noack2011book}}, that the
  reference relation \eqref{Eqn:ReferenceEddyViscosity} is accurate in
  the limit of large $e$.  Thus, Assumption \ref{ass1} ensures that
  for small and large values of the fluctuation energy, for which no
  sensitivity information can be extracted from the model, the optimal
  reconstruction smoothly falls back to the reference eddy viscosity
  \eqref{Eqn:ReferenceEddyViscosity}, or any other relation chosen in
  its place. {We add that from the practical point of view this is not
    a problem, because in any given flow the fluctuation energy will
    not exceed $\max_{t \in [0,T]} E(t)$ by a significant fraction and
    hence the values of $\nu_T^{\bullet}(e)$ for $e > \max_{t \in
      [0,T]} E(t)$ are not very important for the accuracy of the
    reduced-order model (the optimal eddy viscosity is defined for
    such $e$ for technical reasons only).}  It should be also
  emphasized that the optimal eddy viscosity $\nu_T^\bullet(e)$ is
  allowed to become {\em negative} for some values of the turbulent
  kinetic energy $e$.}

The optimization problem for finding $\nu_T^{\bullet}$ can be
therefore stated as follows
\begin{equation}
\nu_T^{\bullet} = \operatorname*{argmin}_{{\stackrel{\nu_T \in H^1(\I),}{\nu_T(0) = 0, \ \frac{d \nu_T}{de}\big|_{e=\Em}=G}}} \ \J(\nu_T) 
\label{Eqn:P}
\end{equation}
with cost functional $\J(\nu_T)$ given in \eqref{Eqn:J} together with
\eqref{Eqn:GalerkinSystem4}. While problem \eqref{Eqn:P} is of the
``parameter identification'' type, it is in fact quite different from
the related problems already studied in the literature on
reduced-order modelling \citep{apma07,accm12,Cordier2013ef}, in which
the optimal eddy viscosity $\nu_T$ was sought as a function of time
(i.e., an independent variable in the problem). The reduced-order
model {resulting from such formulation} is {\em non-autonomous}
and therefore restricted to the time-window and the initial condition
used in the determination of the optimal eddy viscosity. Consequently,
such time-dependent optimal eddy viscosity cannot be considered a
proper ``closure model''. On the other hand, {our} formulation
\eqref{Eqn:P} {is fundamentally different and} leads to an
optimal eddy viscosity as a constitutive relation of the form
$\nu_T^{\bullet}=\nu_T^{\bullet}((1/2)\|\a\|_2^2)$, {so that} the
corresponding reduced-order model is {\em autonomous}.

In order to ensure that optimal eddy viscosity $\nu_T^{\bullet}$
satisfies Assumption \ref{ass1}, we will adopt the
``optimize-then-discretize'' paradigm \citep{g03} in solving problem
\eqref{Eqn:P}. While solution of this problem relies on a standard
gradient-based approach, it requires a specialized technique for the
evaluation of gradients. Its mathematical and computational
foundations were established by \cite{bvp10} and \cite{bp11a}, and
here we use an adaptation of this approach to the identification of
reduced-order models recently developed by \cite{pnm14}. Below we
present the main elements of the algorithm deferring technical details
to Appendix \ref{Sec:Adjoint}.

The (local) minimizer $\nu_T^{\bullet}$ of \eqref{Eqn:J} is
characterized by the first-order optimality condition \citep{l69}
requiring the vanishing of the G\^ateaux differential
$\J'(\nu_T;\nu'_T) := \lim\limits_{\epsilon\rightarrow 0} \epsilon^{-1}
\left[ \J(\nu_T + \epsilon \nu'_T) - \J(\nu_T)\right]$, i.e.,
\begin{equation}
\forall_{\nu'_T \in H^1(\I), \ \nu'_T(0) = 0, \ \frac{d \nu'_T}{de}\big|_{e=\Em}=0 } \qquad \J'(\nu_T^{\bullet} ;\nu'_T) = 0,
\label{Eqn:opt}
\end{equation}
where $\nu'_T$ is an arbitrary perturbation direction. This minimizer
can be computed as $\nu_T^{\bullet} = \lim\limits_{n \rightarrow \infty}
\nu_T^{(n)}$ using the following iterative procedure
\begin{equation}
\left\{
\begin{alignedat}{2}
&\nu_T^{(n+1)} && = \nu_T^{(n)} - \tau^{(n)} \nabla\J(\nu_T^{(n)}),
 \qquad n=1,\dots, \\
&\nu_T^{(1)}   && = \nu_T^{\circ},
\end{alignedat}
\right.
\label{Eqn:desc}
\end{equation}
where the reference eddy viscosity $\nu_T^{\circ}$ from \S\
\ref{Sec:EddyViscosity} is taken as the initial guess, $n$ denotes the
iteration count and $\nabla\J \; : \; \I \rightarrow \RR$ is the
gradient of cost functional $\J$. The length $\tau^{(n)}$ of the step
{along the descent direction} is determined by solving line
minimization problem
\begin{equation}
\tau^{(n)} = \operatorname*{argmin}_{\tau > 0} \; \J\left(\nu_T^{(n)} - \tau \nabla\J_1(\nu_T^{(n)})\right)
\label{eq:taumax}
\end{equation}
which can be done efficiently using standard techniques such as
Brent's method \citep{pftv86}.  For the sake of clarity, formulation
\eqref{Eqn:desc} represents the steepest-descent method, however, in
practice one typically uses more advanced minimization techniques,
such as the conjugate gradient method, or one of the quasi-Newton
techniques \citep{nw00}.  Evidently, the key element of minimization
algorithm \eqref{Eqn:desc} is the computation of the cost functional
gradient $\nabla\J$. It ought to be emphasized that, while the
governing system \eqref{Eqn:GalerkinSystem4} is finite-dimensional,
the gradient $\nabla\J$ is a function of the turbulent kinetic energy
$e$ and as such represents a continuous (infinite-dimensional)
sensitivity of cost functional $\J(\nu_T)$ to the perturbations
$\nu'_T = \nu'_T(e)$. As shown in Appendix \ref{Sec:Adjoint}, the
$L_2$ gradient of cost functional \eqref{Eqn:J} can for $e \in
[0,\Em]$ be evaluated as
\begin{equation}
\nabla^{L_2}\J(e) = \sum_{\stackrel{t}{E(\a(t)) = e}} \frac{\sum_{i,j=0}^N l^{\nu}_{ij} a_j(t) a^*_i(t)}{\sum_{i=1}^N  a_i(t) \left[f_i(\a(t)) + \nu_T((1/2)\| \a(t) \|^2_2)  \sum_{j=0}^N l_{ij}^{\nu} a_j(t)\right]}
\label{Eqn:gradL2}
\end{equation}
in which $f_i(\a(t))$ is defined in \eqref{Eqn:GalerkinSystem},
whereas $\a^*(t) = [0,a^*_1(t), \dots,a^*_N(t)]^T \in \RR^{N+1}$ is
the solution of adjoint system
\begin{subequations}
\label{Eqn:GalerkinAdjoint1}
\begin{align}
 -\frac{da^*_i}{dt} & = \sum_{j=0}^N A_{ji} a^*_j + \frac{a_i}{T} \left[E(t) - \widetilde{E}(t)\right], 
\quad t \in ((m-1)\Delta T, m\Delta T], \label{Eqn:GalerkinAdjoint1a} \\
 a^*_i(m\Delta T) & = 0, \qquad i=1,\dots,N, \quad m=1,\dots,M,
\label{Eqn:GalerkinAdjoint1b}
 \end{align}
\end{subequations}
where $\bA$ is the linearized operator defined in Appendix
\ref{Sec:Adjoint}. So that it has the same dimension $(N+1)$ as the
state vector $\a(t)$, cf.~\S\ \ref{Sec:POD}, the adjoint state $\a^*(t)$
is defined to have an extra (zero) element in the first position.  In
order to ensure that the optimal eddy viscosity $\nu_T^{\bullet}$
possesses the smoothness and boundary behaviour required by Assumption
\ref{ass1}, in iterations \eqref{Eqn:desc} we need to use the $H^1$
Sobolev gradient $\nabla\J = \nabla^{H^1}\J$ defined with respect to
inner product \eqref{Eqn:ipH1}, rather than the $L_2$ gradient given
in \eqref{Eqn:gradL2}. The two gradients are related through the
following elliptic boundary-value problem \citep{pbh04}
\begin{subequations}
\label{Eqn:gradJH1}
\begin{alignat}{2}
\left( 1 - \ell^2 \frac{d^2}{de^2} \right) \nabla^{H^1} \J &= \nabla^{L_2} \J & \qquad & \textrm{in} \ (0,\Em), \label{Eqn:gradJH1_1} \\
\nabla^{H^1} \J &= 0 & & \textrm{at} \ e = 0, \label{Eqn:gradJH1_2} \\
\frac{d}{de} \nabla^{H^1} \J &= 0 & & \textrm{at} \ e = \Em, \label{Eqn:gradJH1_3}
\end{alignat}
\end{subequations}
where $\ell \in \RR$ is a parameter with the meaning of a ``length
scale''. \citet{pbh04} showed that extraction of cost functional
gradients in the space $H^1$ with the inner product defined as in
\eqref{Eqn:ipH1} can be regarded as low-pass filtering the $L_2$
gradients with the cut-off wavenumber given by $\ell^{-1}$.  As
regards the behaviour of the gradients $\nabla^{H^1} \J$ at the
endpoints of the interval $\I$, boundary conditions
\eqref{Eqn:gradJH1_2}--\eqref{Eqn:gradJH1_3} ensure that all iterates
$\nu_T^{(n)}$ have the same behaviour as the initial guess
$\nu_T^{\circ}$, cf.~Assumption \ref{ass1}(c,d). At every iteration
\eqref{Eqn:desc} of the computational algorithm one first evaluates
the $L_2$ gradient \eqref{Eqn:gradL2}, which requires integration
along the system trajectory in the phase space $\RR^N$ \citep{pnm14},
and then solves problem \eqref{Eqn:gradJH1} as a ``post-processing''
step to obtain the Sobolev gradient $\nabla^{H^1}\J$. Application of
this approach to identification of the optimal eddy viscosity in
reduced-order models of two complex flow problems is discussed in the
next section.

\section{Results}
\label{Sec:Results}

In this section we present the results obtained applying the procedure
from \S\ \ref{Sec:OptimalEddyViscosity} to determine the optimal eddy
viscosity $\nu_T^{\bullet}$ for two realistic flow problems with
distinct properties from the point of view of reduced-order modeling.
The first one, discussed in \S\ \ref{Sec:MixingLayer}, concerns a 2D
mixing layer at a medium Reynolds number. It features a small number
of dominating frequencies and most of the flow energy is resolved by a
20-dimensional Galerkin model. The second problem, discussed in \S\
\ref{Sec:AhmedBody}, concerns a high Reynolds number wake flow past an
Ahmed body. This flow problem is characterized by a broadband
frequency spectrum such that a 100-dimensional Galerkin model resolves
less than half of the total energy only.

\subsection{Mixing layer model}
\label{Sec:MixingLayer}

\begin{figure}
\begin{center}
\includegraphics[width=0.8\textwidth]{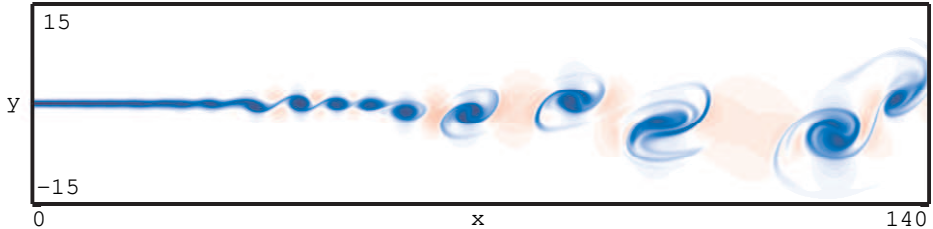}
\caption{An instantaneous vorticity field (with red and blue
  representing, respectively, positive and negative values) in the 2D
  mixing layer flow studied in \S\ \ref{Sec:MixingLayer}.}
\label{fig:ml}
\end{center}
\end{figure}

The 2D mixing layer has a Reynolds number of $500$ based on the
initial vorticity thickness $L=\delta_v$ and the maximum velocity of
the upper stream $U=U_1$.  {The inflow is described by a $\tanh$
  profile with stochastic perturbations and} the velocity ratio
between the upper and lower stream is {equal to} $U_1/U_2 = 3$.
The observation region for the POD analysis {coincides with} the
computational domain and {is} given by
\begin{equation}
\Omega := \left \{ (x,y) \colon  0 \le x \le 140, \quad -28 \le y \le 28 \right \}.
\end{equation}
{The DNS is based on the 6th-order accurate compact
  finite-difference approximations for the derivatives in space and a
  3rd-order accurate approximation for the derivatives with respect to
  time.}  The post-transient flow is computed over 2000 convective
time units and sampled with the uniform time step $\Delta t = 1$.
{Further details concerning the numerical approach} are described
by \citet{Kasten2014arXiv,Kaiser2014jfm}, and figure \ref{fig:ml}
shows a snapshot of the vorticity field in the flow. The numerical
data is used to construct Galerkin system
\eqref{Eqn:GalerkinSystemEddyViscosity} with dimension $N=20$ using
the procedure discussed in \S\ \ref{Sec:ROM} and setting $\kappa_i =
1$, $i=1,\dots,N$, in \eqref{Eqn:ReferenceEddyViscosity}. The
dimension $N=20$ ensures that the Galerkin system captures $80\%$ of
the flow energy.  Optimization problem \eqref{Eqn:P} is solved for a
broad range of time intervals $4 \le \Delta T \le 2000$ ($500 \ge M
\ge 1$) at which the governing system \eqref{Eqn:GalerkinSystem4} is
restarted with new initial conditions.  {Generally, optimal eddy
  viscosities with two distinct sets of properties are recovered and
  in order to illustrate these reconstructions} below we will present
the results for two representative cases with $\Delta T = 10$ and
$\Delta T = 200$ which will be referred to as optimization over,
respectively, short and long windows.

\begin{figure}
\begin{center}
\subfigure[]{\includegraphics[width=0.8\textwidth]{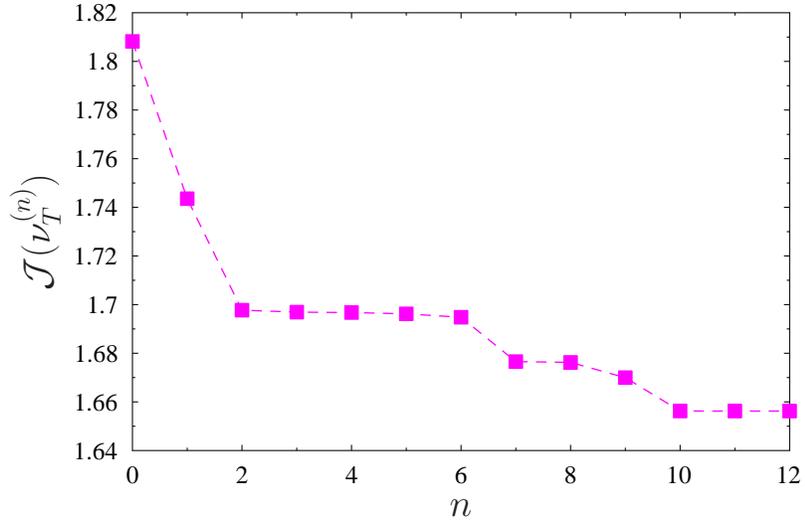}}
\subfigure[]{\includegraphics[width=0.8\textwidth]{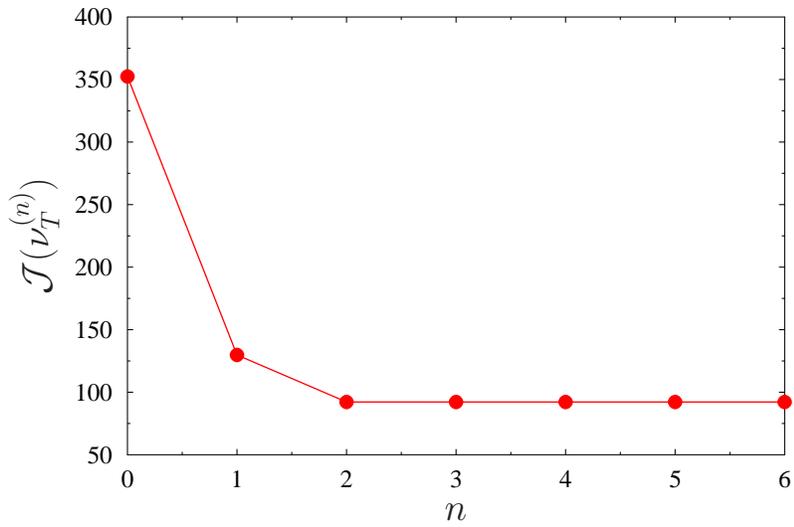}}
\caption{[Mixing layer] Decrease of the cost functional \eqref{Eqn:J}
  with iterations $n$ for optimization over (a) over short windows
  ($\Delta T = 10$) and (b) over long windows ($\Delta T = 200$).
  {The two sets of data are plotted on separate graphs because of
    the widely different values of $\J(\nu_T^{(n)})$.} }
\label{fig:Jml}
\end{center}
\end{figure}

\begin{figure}
\begin{center}\includegraphics[width=0.8\textwidth]{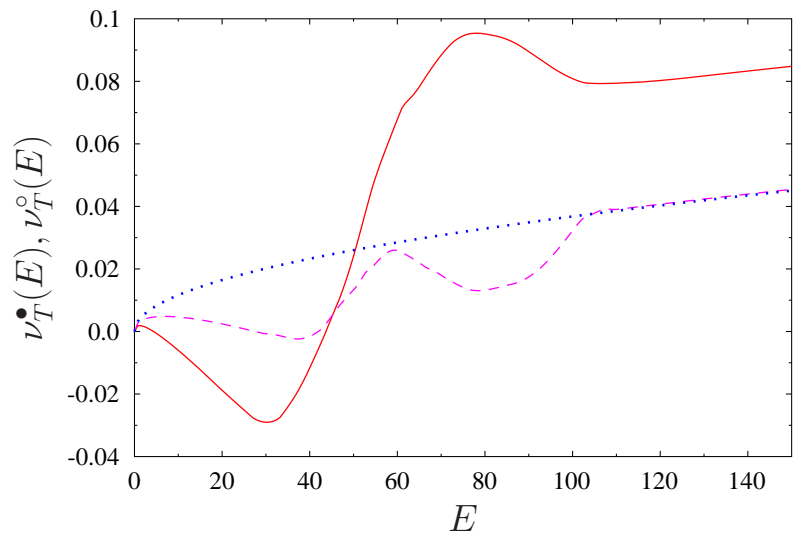}
  \caption{[Mixing layer] Optimal eddy viscosity $\nu_T^{\bullet}(E)$
    obtained with optimization over long windows ($\Delta T = 200$;
    red solid line) and over short windows ($\Delta T = 10$; dashed
    purple line); {reference} eddy viscosity $\nu_T^{\circ}(E)$
    is marked thick blue dotted line.}
\label{fig:nuTml}
\end{center}
\end{figure}

\begin{figure}
\begin{center}
\subfigure[]{\includegraphics[width=1.0\textwidth]{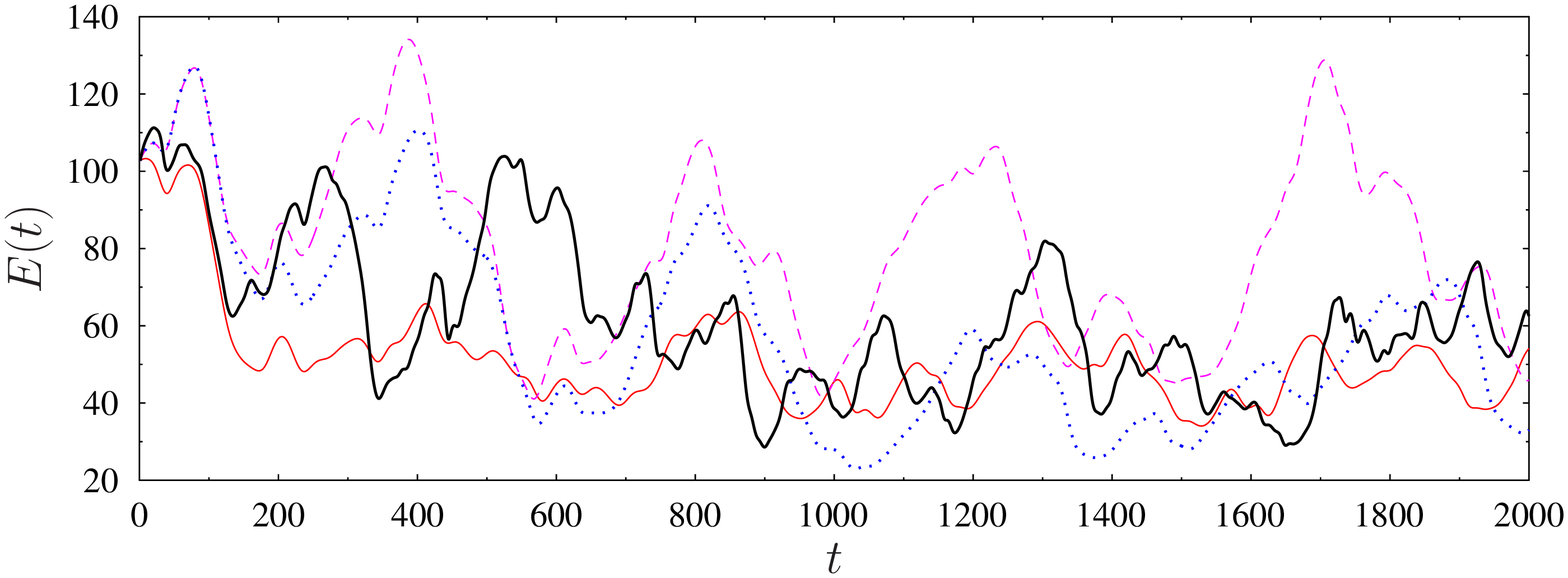}}
\subfigure[]{\includegraphics[width=1.0\textwidth]{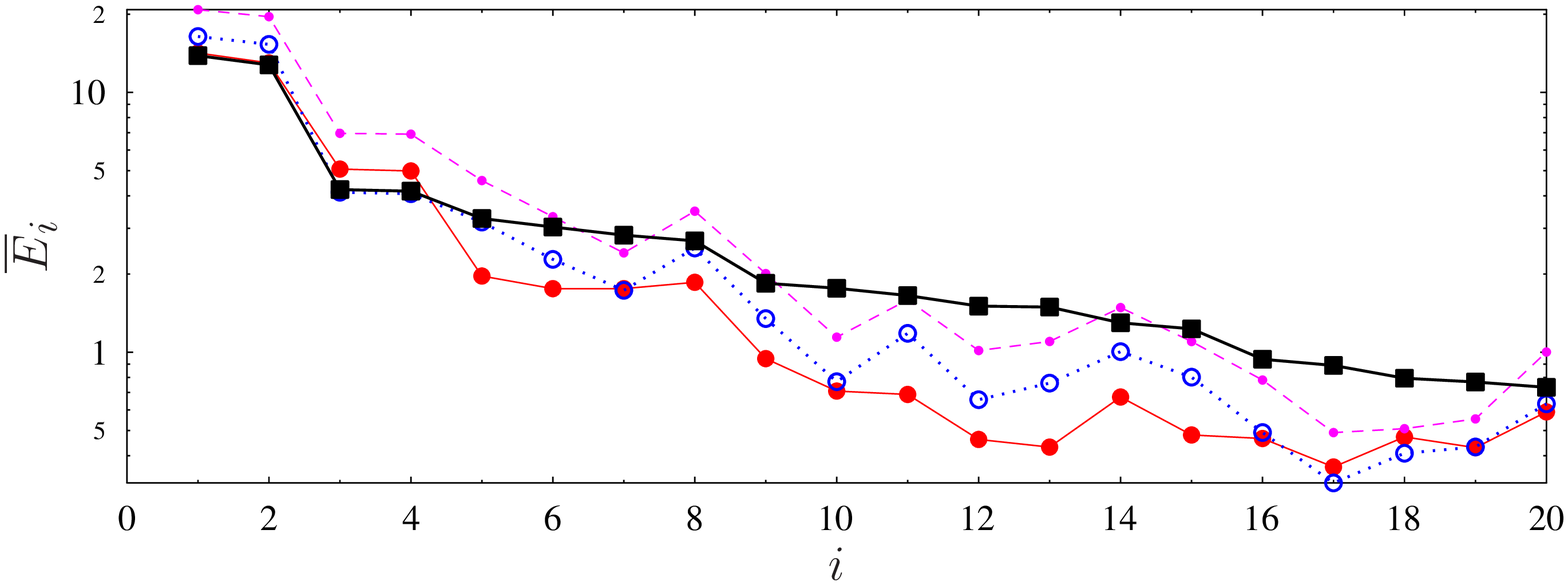}}
\caption{[Mixing layer] (a) Turbulent kinetic energy $E(t)$ as a
  function of time $t$ and (b) time-averaged modal energy
  $\overline{E}_i$ as a function of mode index $i$ for DNS projected
  on $N=20$ POD modes (thick black solid line), ROM with the
  {reference} eddy viscosity $\nu_T^{\circ}(E)$ (thick blue
  dotted line) and the optimal eddy viscosity $\nu_T^{\bullet}(E)$
  obtained with optimization over long windows ($\Delta T = 200$; red
  solid line) and over short windows ($\Delta T = 10$; dashed purple
  line).}
\label{fig:Ktml}
\end{center}
\end{figure}

We begin by presenting in figure \ref{fig:Jml} the decrease of cost
functional \eqref{Eqn:J} with iterations \eqref{Eqn:desc}. We see that
in the case of the short window ($\Delta T = 10$) not only are the
values of functional \eqref{Eqn:J} smaller, but also the relative
decrease achieved during iterations is less significant (about $8\%$
in figure \ref{fig:Jml}a). This implies that over such short time
windows the {reference} ansatz \eqref{Eqn:ReferenceEddyViscosity}
for eddy viscosity performs satisfactorily and the improvement
obtained with optimization is marginal only. On the other hand, in the
case with longer time windows ($\Delta T = 200$, see figure
\ref{fig:Jml}b), the values of the cost functional are much larger as
is its relative reduction (about $74\%$) achieved with optimization.
The corresponding optimal eddy viscosities $\nu_T^{\bullet}$ are
presented in figure \ref{fig:nuTml} together with the {reference}
relation \eqref{Eqn:ReferenceEddyViscosity}. {We see that the
  optimal relation $\nu_T^{\bullet}$ deviates from the reference eddy
  viscosity $\nu_T^{\circ}$ for} $E \in [0,100]$, which is the range
of values spanned by the DNS solution, see figure \ref{fig:Ktml}a.
{On the other hand,} for values of $E$ outside that range the
{sensitivity information is not available and therefore by
  construction, cf.~Assumption \ref{ass1}(d), the optimal eddy
  viscosity $\nu_T^{\bullet}$ exhibits the same behaviour as the
  reference relation $\nu_T^{\circ}$.}  Two distinct behaviours are
observed, with the optimal eddy viscosity $\nu_T^{\bullet}$ becoming
negative for $E\in [0,40]$ in the case with optimization over long
windows ($\Delta T = 200$). We remark that this feature of the eddy
viscosity was already discussed in \S\ \ref{Sec:TransientTimes} where
it was found to arise in a two-dimensional Galerkin model of laminar
vortex shedding in the cylinder wake. {The bimodal form of the
  optimal eddy viscosity shown in figure \ref{fig:nuTml} for the short
  optimization window helps stabilize multiple energy levels in the
  flow. On the other hand, the negative eddy viscosity obtained with
  long optimization windows creates an excitation mechanism for the
  coherent structures. The physical aspects of the optimal eddy
  viscosities are further discussed and compared among different flow
  problems in \S\ \ref{Sec:Conclusions}.}

\begin{figure}
\begin{center}
\includegraphics[width=1.0\textwidth]{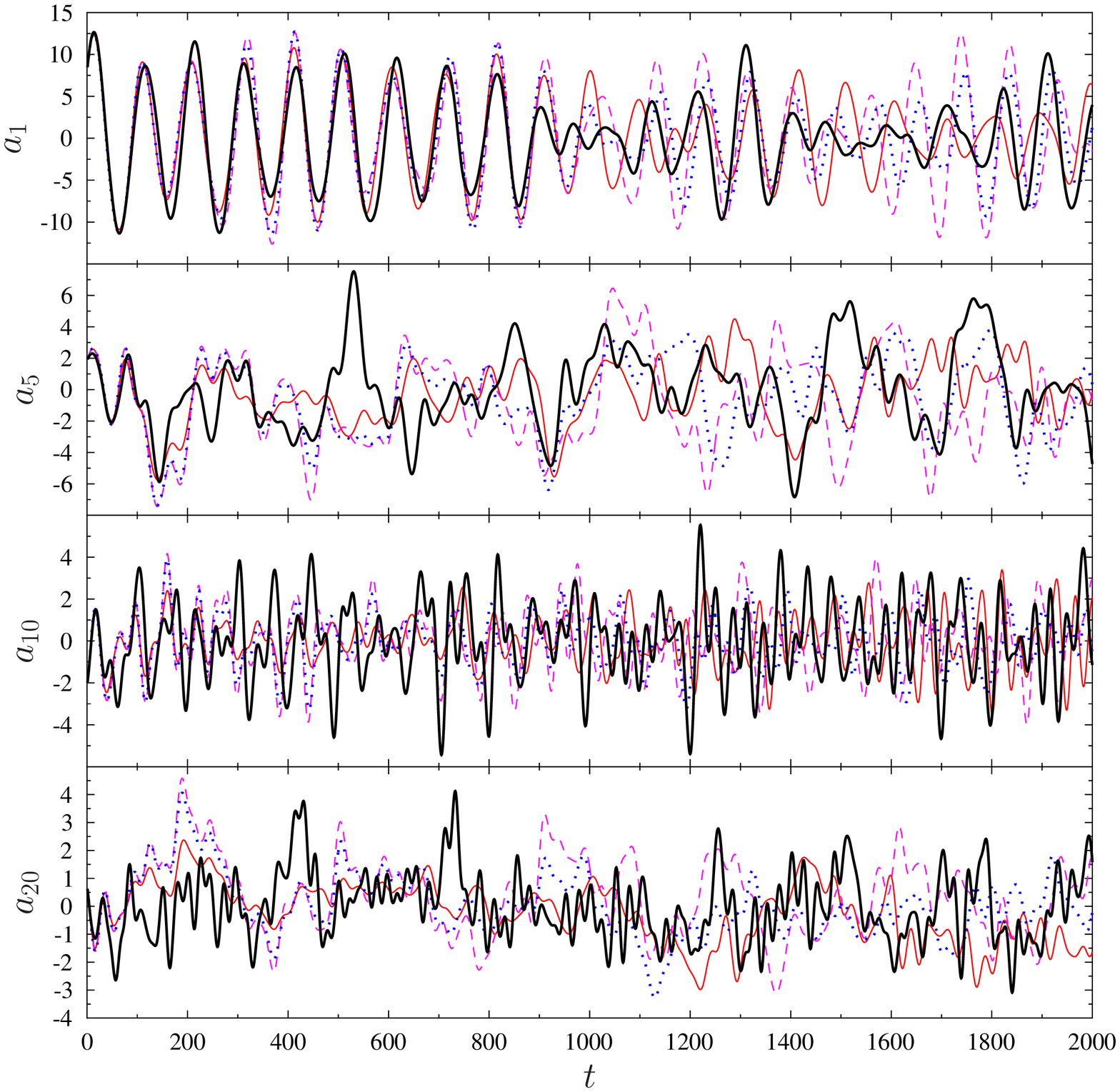}
\caption{[Mixing layer] Galerkin expansion coefficients $a_k(t)$,
  $k=1,5,10,20$, as a function of time $t$ for DNS projected on $N=20$
  POD modes (thick black solid line), ROM with the {reference}
  eddy viscosity $\nu_T^{\circ}(E)$ (thick blue dotted line) and the
  optimal eddy viscosity $\nu_T^{\bullet}(E)$ obtained with
  optimization over long windows ($\Delta T = 200$; red solid line)
  and over short windows ($\Delta T = 10$; dashed purple line).}
\label{fig:aiml}
\end{center}
\end{figure}

\begin{figure}
\begin{center}
\includegraphics[width=1.0\textwidth]{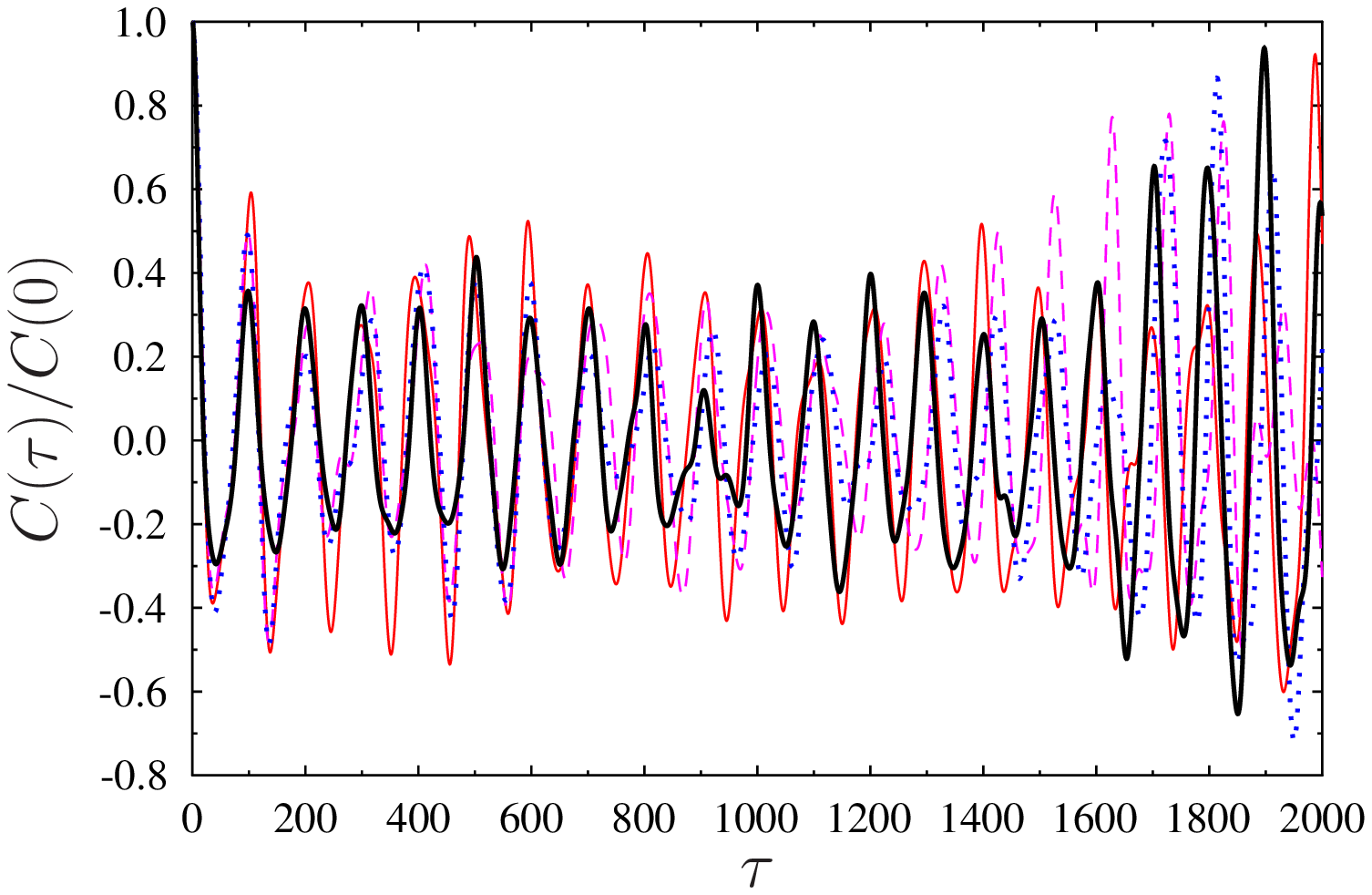}
\caption{[Mixing layer] Normalized unbiased two-time correlation
  function $C(\tau)/C(0)$ for DNS projected on $N=20$ POD modes (thick
  black solid line), ROM with the {reference} eddy viscosity
  $\nu_T^{\circ}(E)$ (thick blue dotted line) and the optimal eddy
  viscosity $\nu_T^{\bullet}(E)$ obtained with optimization over long
  windows ($\Delta T = 200$; red solid line) and over short windows
  ($\Delta T = 10$; dashed purple line).}
\label{fig:corrml}
\end{center}
\end{figure}

The histories of the resolved total kinetic energy $E(t)$, which is
the quantity used as the performance criterion in our optimization
problem, cf.~\eqref{Eqn:J}, are presented in figure \ref{fig:Ktml}a,
whereas in figure \ref{fig:Ktml}b we show the corresponding average
modal energies $\overline{E}_i$, $i=1,\dots,20$.  The mean values of
the total kinetic energy $\overline{E}$ and their standard deviations
are summarized in Table \ref{tab:ml}. {An interesting observation one
can make about this data is that the standard deviation of the
turbulent kinetic energy is quite high and equal to about a third of
its mean value $\overline{E}$. The reason is that the mixing-layer
flow is dominated by a relatively small number of large coherent
structures (cf.~figure \ref{fig:ml}). Although this may not be evident
from the data in Table \ref{tab:ml}, figure \ref{fig:Ktml}a shows that
the optimal eddy viscosity $\nu_T^{\bullet}$ obtained with
optimization over long windows ($\Delta T = 200$) allows Galerkin
system \eqref{Eqn:GalerkinSystemEddyViscosity} to track the total
kinetic energy $\widetilde{E}(t)$ of the original DNS simulation
better than when the reference ansatz $\nu_T^{\circ}$ is used. This
improvement is quantified by {a} $74\%$ decrease of the cost
functional, representing the least-squares reconstruction error,
cf.~\eqref{Eqn:J}, starting from the initial guess given by the
reference relation $\nu_T^{\circ}$ and the final iteration producing
the optimal reconstruction $\nu_T^{\bullet}$ (figure \ref{fig:Jml}b).
Figure \ref{fig:Ktml}b indicates that this improvement is achieved
with the optimal eddy viscosity $\nu_T^{\bullet}$ by a more accurate
reconstruction of the average modal energy of the first two modes
which comes at the price of a somewhat poorer reconstruction of $E_i$
when $i\ge 2$.}  On the other hand, when the optimal eddy viscosity is
obtained with optimization over short windows ($\Delta T = 10$), only
a modest improvement is observed.  {The reason for that is that, as
will be discussed in more detail in \S\ \ref{Sec:Conclusions}, the
optimization horizon $\Delta T = 10$ is shorter than the time scale of
the characteristic events in the flow.}  These observations are also
corroborated by the results presented in figure \ref{fig:aiml}, where
we show the time-histories of selected Galerkin coefficients $a_i(t)$,
$i=1,5,10,20$.  {In that figure we see that the optimal eddy viscosity
$\nu_T^{\bullet}$ obtained with long optimization windows allows one
to capture the amplitude $a_1$ of the first POD mode with a higher
accuracy than when the reference relation $\nu_T^{\circ}$ is used. On
the other hand, this optimal eddy viscosity tends to underestimate the
amplitudes of the higher modes with $i=5,10,20$.  Such trade-offs,
which are typical of solutions obtained with optimization approaches,
are a consequence of our choice of the cost functional \eqref{Eqn:J}
based on energy, a quantity which in the present flow is captured by
the first few POD modes (figure \ref{fig:Ktml}b). In other words, POD
modes with $i\ge 3$ contribute much less to the cost functional than
the first two modes, and therefore their behaviour is to a lesser
extent improved by optimization.} In figure \ref{fig:corrml} we present
the ``unbiased'' correlation function \citep{o96}
\begin{equation}
C(\tau) := \frac{1}{T-\tau}\int_{\tau}^T \langle \boldsymbol{u}'(\cdot,t-\tau)\cdot \boldsymbol{u}'(\cdot,t)\rangle_{L_2(\Omega)}\, dt, \quad \tau \in [0,T)
\label{Eqn:Corr}
\end{equation}
after normalization with respect to $C(0)$.  We note that using ansatz
\eqref{Eqn:POD+Residual} and the orthogonality property of the POD
modes, it can be conveniently evaluated in terms of the
autocorrelations of the individual Galerkin coefficients, i.e.,
\begin{equation}
C(\tau) = \frac{1}{T-\tau} \sum_{i=1}^N \int_{\tau}^T a_i(t-\tau) a_i(t) \, dt.
\label{Eqn:Corr2}
\end{equation}
{In figure \ref{fig:corrml} illustrating this correlation function the
oscillatory behaviour at levels around 0.3 reveals a dominant
periodicity in the mixing layer. This rather low level comes from the
fact that any vortex configuration is a new realization and is never
exactly reproduced at any other time. The increasing correlation level
as $\tau \to 2000$ indicates that the final state is close to the
initial one. The large numerical values result from the narrowing of
the integration window in \eqref{Eqn:Corr2} and the corresponding
normalization. Due to this effect, there is hardly any averaging
possible for large values of the correlation time $\tau$.}

\begin{figure}
\begin{center}
\subfigure[]{\includegraphics[width=1.0\textwidth]{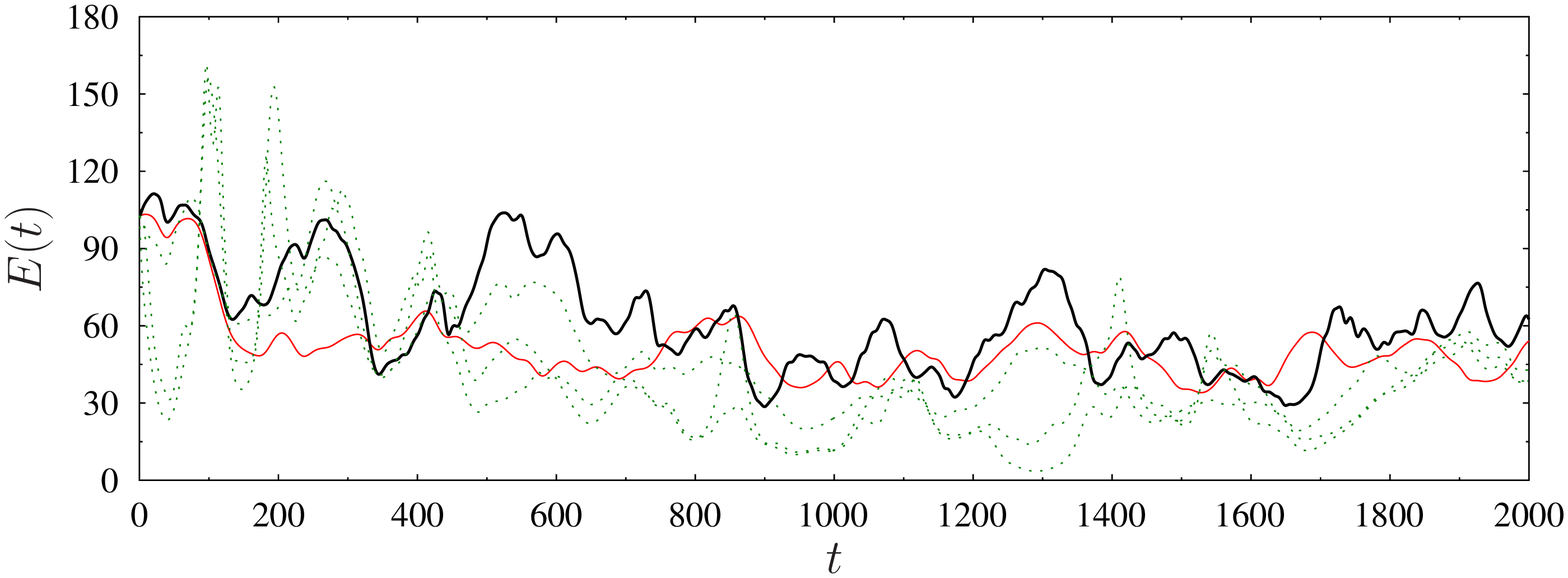}}
\caption{[Mixing layer] Comparison of turbulent kinetic energy $E(t)$
  as a function of time $t$ for DNS (thick black solid line), optimal
  reconstruction on long windows ($\Delta T = 200$; thin red solid
  line) and the results from \cite{Cordier2013ef} corresponding to
  three different objective functionals (dotted green lines).}
\label{fig:Ktml2}
\end{center}
\end{figure}

Finally, in figure \ref{fig:Ktml2} we compare our results concerning
the history of the total kinetic energy $E(t)$ with the results
obtained by \cite{Cordier2013ef} who used an optimization approach to
determine eddy viscosities as functions of time $\nu_T = \nu_T(t)$
with different cost functionals. We see that the optimization
formulation proposed here, in which the optimal eddy viscosity is
sought as a function of the instantaneous turbulent kinetic energy
$\nu_T^{\bullet} = \nu_T^{\bullet}(E)$, leads to {a more accurate
  tracking of the energy $\widetilde{E}(t)$ characterizing the DNS
  than any of the time-dependent eddy viscosities $\nu_T(t)$,
  especially at later times ($t > 800$).}

\begin{table}
\begin{center}
  \caption{[Mixing layer] Mean resolved turbulent kinetic energy
    $\overline{E}$ and its standard deviation $\mathop{std}(E)$ in the
    different cases considered in \S\ \ref{Sec:MixingLayer}.}
\vspace{1pc}
\begin{tabular}{ l | c | c | c | c } 
   & \Bmp{2.25cm}\ \Emp&\Bmp{2.25cm}\ \Emp & \Bmp{3.5cm}\Emp & \Bmp{3.5cm}\Emp \\
   & Original DNS 
   & System \eqref{Eqn:GalerkinSystemEddyViscosity}
   & System \eqref{Eqn:GalerkinSystemEddyViscosity} with $\nu_T^{\bullet}$ 
   & System \eqref{Eqn:GalerkinSystemEddyViscosity} with $\nu_T^{\bullet}$  
\\ & ($N=20$) 
   &  with $\nu_T^{\circ}$  
   & short windows
   &  long windows 
\\ & &
   &  ($\Delta T = 10$) 
   &  ($\Delta T = 200$) \rule[0pt]{0pt}{0pt} 
\\ & \Bmp{2.25cm}\ \Emp&\Bmp{2.25cm}\ \Emp & \Bmp{3.5cm}\Emp & \Bmp{3.5cm}\Emp 
\\ \cline{1-5}
$\overline{E}$ & 61.73 &  58.43 & 80.84 & 51.27 \rule[-10pt]{0pt}{25pt} \\ 
$\mathop{std}(E)$ & 20.12 & 24.79 &  23.77 & 13.41 \rule[-10pt]{0pt}{20pt} 
\label{tab:ml}
\end{tabular}
\end{center}
\end{table}

\clearpage

\subsection{Ahmed body wake model}
\label{Sec:AhmedBody}

\begin{figure}
\begin{center}
\includegraphics[width=0.8\textwidth]{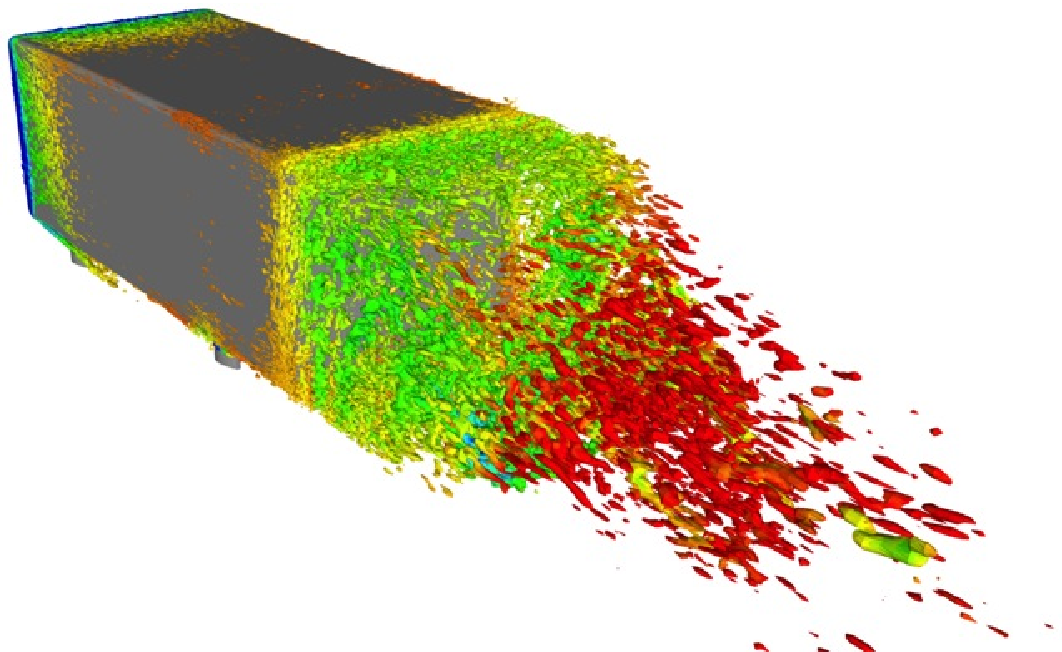}
\caption{Illustration of a typical flow pattern in the turbulent wake
  behind an Ahmed body \citep{Osth2014jfm}. The flow is visualized
  using the quantity $Q(\boldsymbol{x},t) := \bomega\cdot\bomega - \bS :
  \bS$ where $\bomega := \bnabla\times\boldsymbol{u}$ is the vorticity
  and $\bS := (1/2)\left[ \bnabla \boldsymbol{u} + (\bnabla
    \boldsymbol{u})^T\right]$ is the symmetric part of the velocity
  gradient tensor.}
\label{fig:ab}
\end{center}
\end{figure}

The 3D flow over the blunt Ahmed body has the Reynolds number
$Re=300,000$ based on the height $L=H$ of the body and the oncoming
velocity $U=U_{\infty}$.  {The computational domain has
  dimensions $28H \times 8.05H \times 5.35 H$ (length $\times$ width
  $\times$ height), whereas} the observation domain is a small
wake-centered subset of the computational domain:
\begin{equation}
\label{eq:reduceddomain}
{\Omega_0} := \left\{ 
(x,y,z) \in \Omega \> \colon \> 0\le x \le 5\,H, -0.67\,H \le y \le 1.12\,H, \vert z \vert \le 1.21\,H \right\}.
\end{equation} 
This domain is large enough to resolve the recirculation region and
the absolutely unstable wake dynamics, but at the same time small
enough to keep the model dimension affordable. {The LES equations are
discretized in space using a hybrid of central differencing and upwind
schemes applied to the convective fluxes and second-order central
differences applied to the viscous and subgrid terms. The
time-discretization is performed with a second-order accurate implicit
method. {A} computational grid consisting of approximately 34
million mesh points ensures that the LES is well resolved.}  The
post-transient flow is computed over 250 convective time units, which
is half of the time window analyzed by \citet{Osth2014jfm}, and
sampled with the uniform time step $\Delta t=1$.  The reason for
taking a shorter time window is that optimization problem
\eqref{Eqn:P} becomes hard to solve for very large $T$.
{Further} details of the large eddy simulation are described by
\citet{Osth2014jfm} and a typical flow pattern is illustrated in
figure \ref{fig:ab}.  {As expected from a flow at this Reynolds number,
this flow pattern exhibits highly complex multiscale vortex
structures, which makes it quite different from the mixing-layer flow
illustrated in figure \ref{fig:ml}.}  The numerical data is used to
construct Galerkin system \eqref{Eqn:GalerkinSystemEddyViscosity} with
dimension $N=100$ using the procedure discussed in \S\ \ref{Sec:ROM}.
In contrast to the example studied in \S\ \ref{Sec:MixingLayer}, in
the present problem with the chosen dimension $N=100$ the Galerkin
model captures only about $35\%$ of the turbulent kinetic energy of
the entire flow.  We emphasize that the ``target'' turbulent kinetic
energy $\widetilde{E}(t)$ is computed based on the projection of the
actual flow evolution on the first $N=100$ modes, rather than based on
the entire flow field. As in the case of the mixing layer, {we
  performed optimization calculations for a range of different $\Delta
  T$ and below} we will show {the} results {corresponding to} two
{representative} time intervals, namely, $\Delta T=20$ and $\Delta T =
200$, which will be referred to as the short and long window,
respectively.
 
\begin{figure}
\begin{center}
\includegraphics[width=0.8\textwidth]{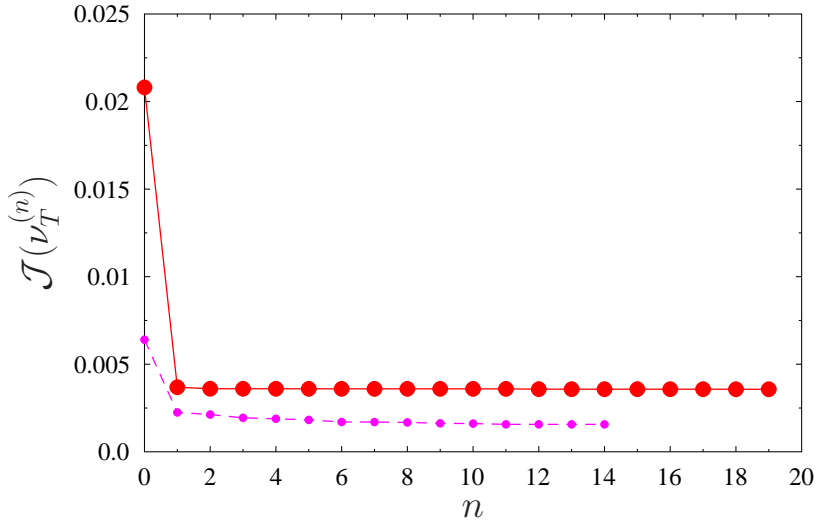}
\caption{[Ahmed body] Decrease of the cost functional \eqref{Eqn:J}
  with iterations $n$ for optimization over short windows ($\Delta T =
  20$; small purple symbols) and over long windows ($\Delta T = 200$;
  big red symbols).  }
\label{fig:Jab}
\end{center}
\end{figure}

\begin{table}
\begin{center}
  \caption{[Ahmed body] Mean resolved turbulent kinetic energy
    $\overline{E}$ and its standard deviation $\mathop{std}(E)$ in the
    different cases considered in \S\ \ref{Sec:AhmedBody}.}
\vspace{1pc}
\begin{tabular}{ l | c | c | c | c } 
   & \Bmp{2.25cm}\ \Emp&\Bmp{2.25cm}\ \Emp & \Bmp{3.5cm}\Emp & \Bmp{3.5cm}\Emp \\
   & Original LES 
   & System \eqref{Eqn:GalerkinSystemEddyViscosity}
   & System \eqref{Eqn:GalerkinSystemEddyViscosity} with $\nu_T^{\bullet}$ 
   & System \eqref{Eqn:GalerkinSystemEddyViscosity} with $\nu_T^{\bullet}$  
\\ & ($N=100$) 
   &  with $\nu_T^{\circ}$  
   & short windows
   &  long windows 
\\ & &
   &  ($\Delta T = 20$) 
   &  ($\Delta T = 200$) \rule[0pt]{0pt}{0pt} 
\\ & \Bmp{2.25cm}\ \Emp&\Bmp{2.25cm}\ \Emp & \Bmp{3.5cm}\Emp & \Bmp{3.5cm}\Emp 
\\ \cline{1-5}
$\overline{E}$ & 0.2739 & 0.4514  & 0.3315 & 0.2759 \rule[-10pt]{0pt}{25pt} \\ 
$\mathop{std}(E)$ & 0.0820  & 0.0562 & 0.0504 & 0.0570  \rule[-10pt]{0pt}{20pt} 
\label{tab:ab}
\end{tabular}
\end{center}
\end{table}

\begin{figure}
\begin{center}
\includegraphics[width=0.8\textwidth]{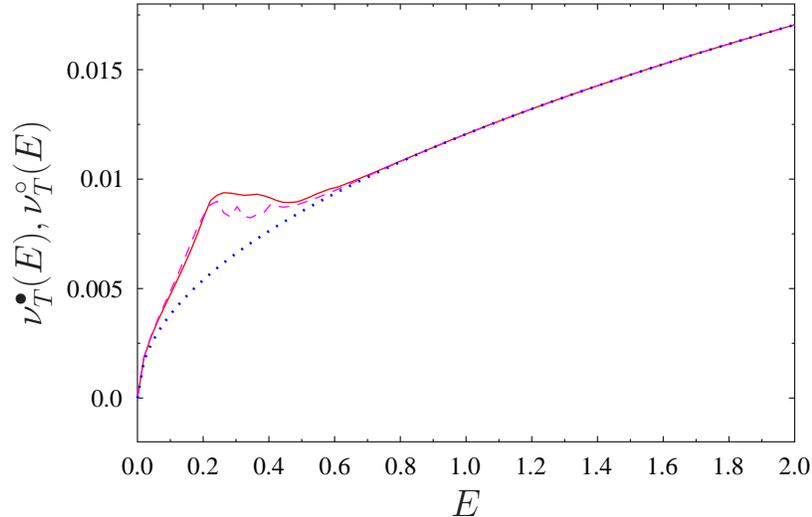}
\caption{[Ahmed body] Optimal eddy viscosity $\nu_T^{\bullet}(E)$
  obtained with optimization over long windows ($\Delta T = 200$; red
  solid line) and over short windows ($\Delta T = 20$; dashed purple
  line); reference eddy viscosity $\nu_T^{\circ}(E)$ is marked thick
  blue dotted line.}
\label{fig:nuTab}
\end{center}
\end{figure}

\begin{figure}
\begin{center}
\subfigure[]{\includegraphics[width=1.0\textwidth]{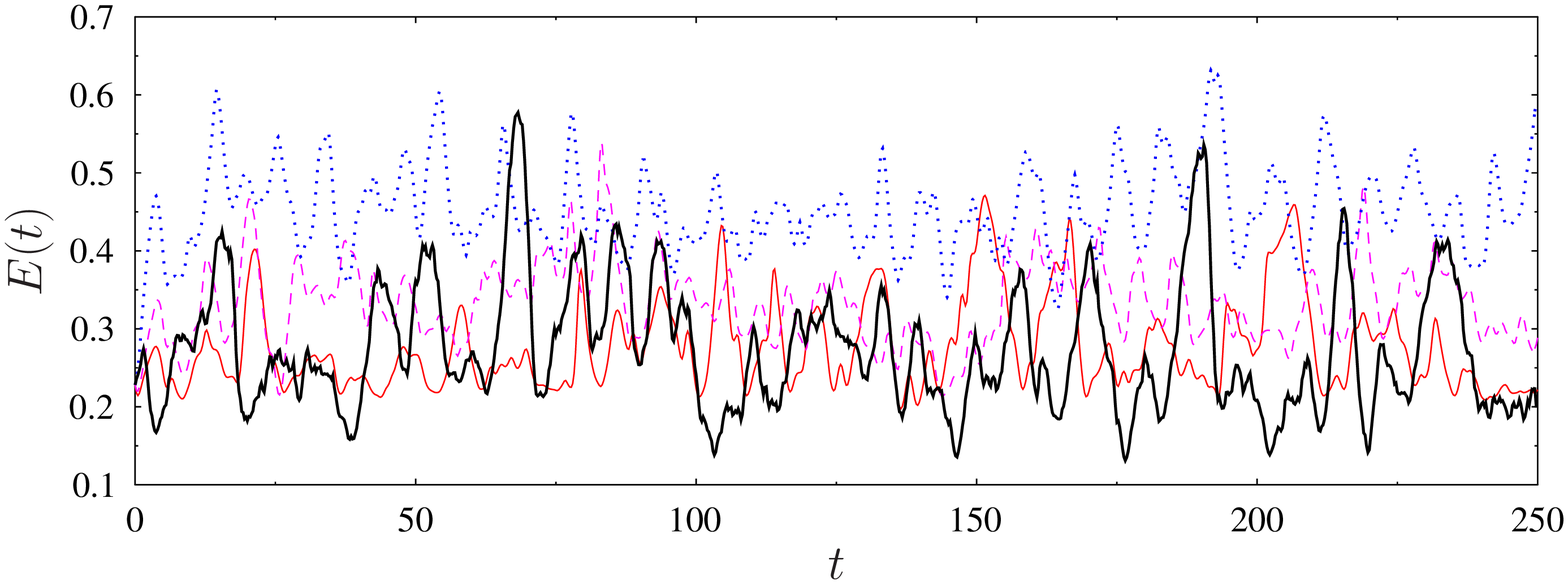}}
\subfigure[]{\includegraphics[width=1.0\textwidth]{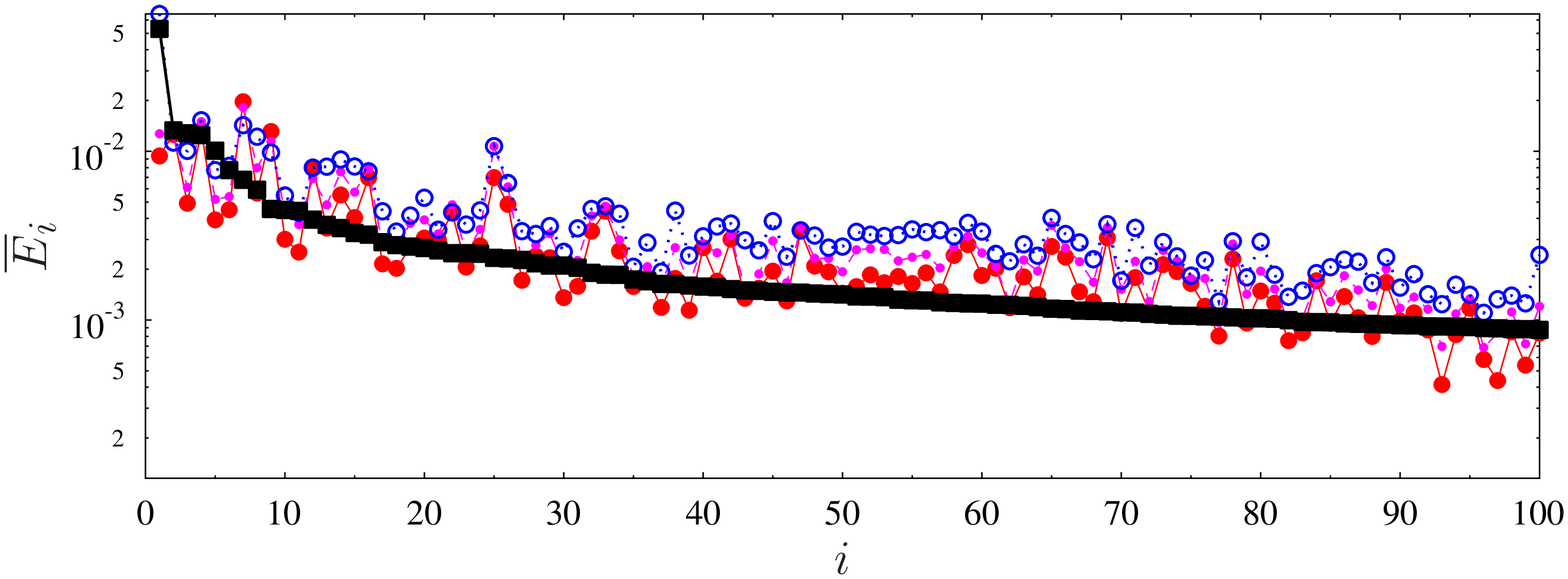}}
\caption{[Ahmed body] (a) Turbulent kinetic energy $E(t)$ as a
  function of time $t$ and (b) time-averaged modal energy
  $\overline{E}_i$ as a function of mode index $i$ for LES projected
  on $N=100$ POD modes (thick black solid line), ROM with the
  reference eddy viscosity $\nu_T^{\circ}(E)$ (thick blue dotted
  line) and the optimal eddy viscosity $\nu_T^{\bullet}(E)$ obtained
  with optimization over long windows ($\Delta T = 200$; red solid
  line) and over short windows ($\Delta T = 20$; dashed purple line).}
\label{fig:Ktab}
\end{center}
\end{figure}

\begin{figure}
\begin{center}
\includegraphics[width=1.0\textwidth]{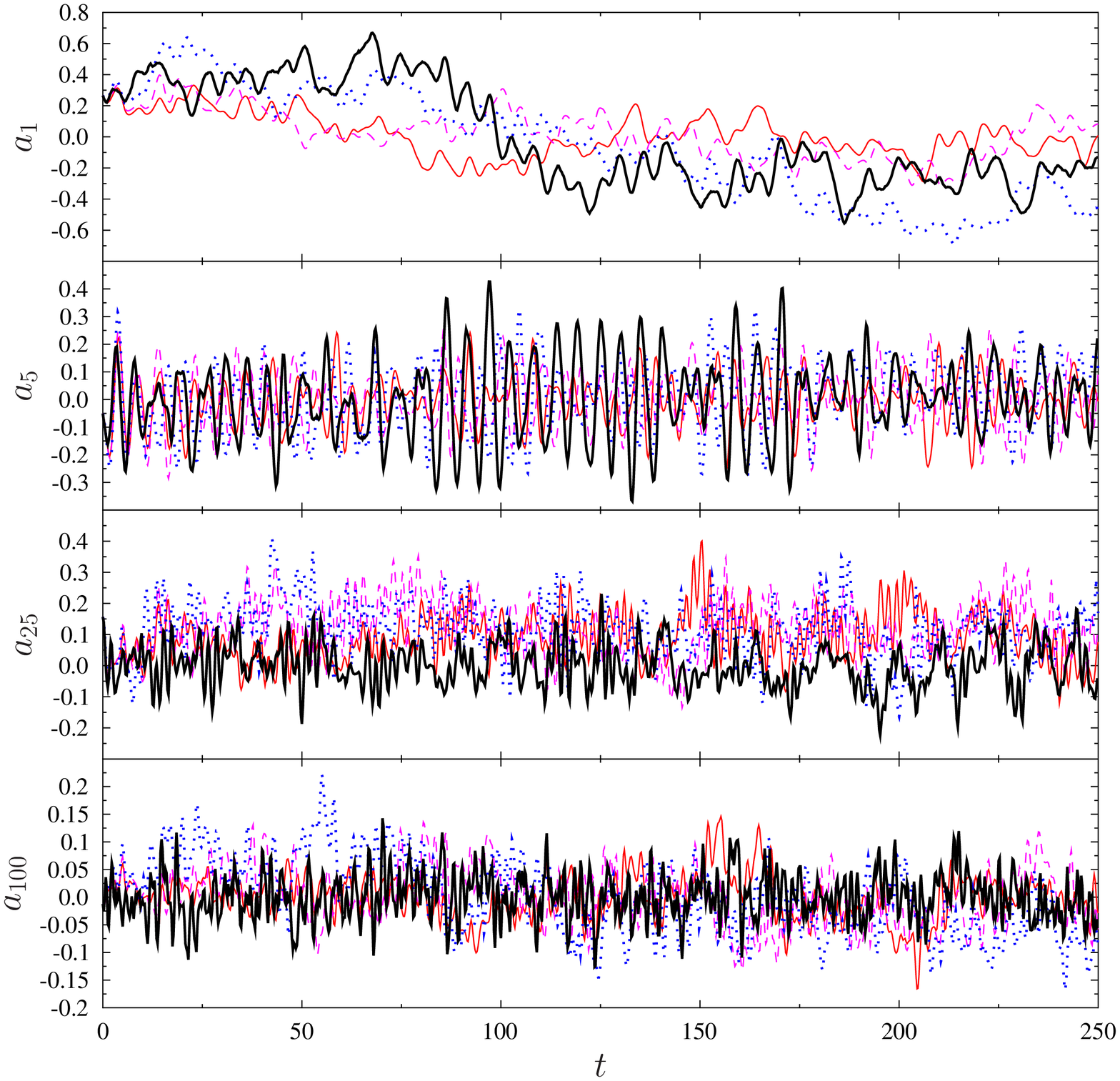}
\caption{[Ahmed body] Galerkin expansion coefficients $a_k(t)$,
  $k=1,5,25,100$, as a function of time $t$ for LES projected on
  $N=100$ POD modes (thick black solid line), ROM with the reference
  eddy viscosity $\nu_T^{\circ}(E)$ (thick blue dotted line) and the
  optimal eddy viscosity $\nu_T^{\bullet}(E)$ obtained with
  optimization over long windows ($\Delta T = 200$; red solid line)
  and over short windows ($\Delta T = 20$; dashed purple line).}
\label{fig:aiab}
\end{center}
\end{figure}

\begin{figure}
\begin{center}
\includegraphics[width=1.0\textwidth]{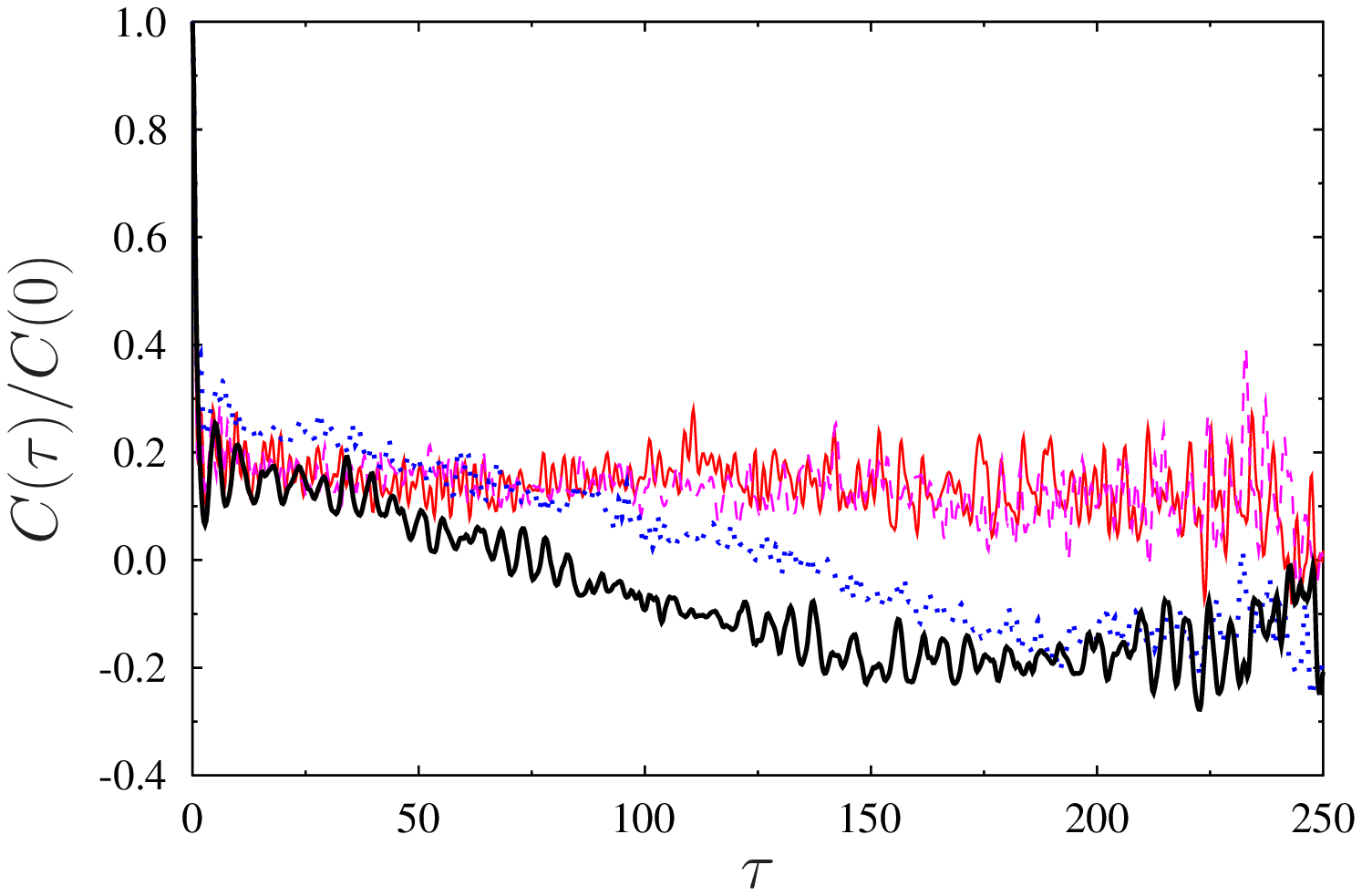}
\caption{[Ahmed body] Normalized unbiased two-time correlation
  function $C(\tau)/C(0)$ for LES projected on $N=100$ POD modes
  (thick black solid line), ROM with the reference eddy viscosity
  $\nu_T^{\circ}(E)$ (thick blue dotted line) and the optimal eddy
  viscosity $\nu_T^{\bullet}(E)$ obtained with optimization over long
  windows ($\Delta T = 200$; red solid line) and over short windows
  ($\Delta T = 20$; dashed purple line).}
\label{fig:corrab}
\end{center}
\end{figure}

Decrease of cost functional \eqref{Eqn:J} with iterations is shown in
figure \ref{fig:Jab} in which a significant reduction can be observed
in both cases. {This implies that the reference eddy viscosity
  \eqref{Eqn:ReferenceEddyViscosity} can be improved by performing
  optimization on either short or long time intervals. The values of
  the cost functional corresponding to long optimization intervals are
  again larger which is due to the fact that, with fewer restarts, the
  trajectory of \eqref{Eqn:GalerkinSystem4} diverges further away from
  the projected trajectory of the actual flow.}  The resulting optimal
eddy viscosities $\nu_T^{\bullet}$ are presented in figure
\ref{fig:nuTab}, together with the {reference relation}
\eqref{Eqn:ReferenceEddyViscosity}. We see that the obtained profile
of the optimal eddy viscosity has a similar general form for both
values of $\Delta T$, except that it is smoother for the case of the
longer window. This suggests that allowing for a longer assimilation
interval before the constraint system \eqref{Eqn:GalerkinSystem4} is
restarted with a new initial condition may have a regularizing effect
{(i.e., may produce smoother optimal eddy viscosity relations)}.
We also {note} that, in contrast to the findings of \S\
\ref{Sec:MixingLayer}, in the present case the optimal eddy viscosity
$\nu_T^{\bullet}$ is uniformly increased with respect to the reference
relation $\nu_T^{\circ}$. {While the function
  $\nu_T^{\bullet}(E)$ is defined for $E \in [0,2]$, {cf.~Assumption
    \ref{ass1}(a)}, deviations from the reference relation
  $\nu_T^{\circ}(E)$ are confined to the range $[0,0.7]$ which is
  {approximately} the range of energy values visited by the system
  trajectory, {(more precisely, $\max_{t \in [0,T]}E(t) \approx 0.6$
    as can be seen from figure \ref{fig:Ktab}a)}. Outside that range
  the sensitivity information is not available and the optimal eddy
  viscosity $\nu_T^{\bullet}$ essentially coincides with the reference
  relation $\nu_T^{\circ}$, cf.~Assumption \ref{ass1}(d). Discussion
  of the physical aspects of the optimal eddy viscosities obtained for
  the Ahmed body wake is deferred to \S\ \ref{Sec:Conclusions}.}
Figure \ref{fig:Ktab}a shows the improvement in the tracking of the
instantaneous turbulent kinetic energy $\widetilde{E}(t)$ achieved by
Galerkin system \eqref{Eqn:GalerkinSystemEddyViscosity} with the
optimal eddy viscosity $\nu_T^{\bullet}$ with respect to the use of
the reference relation $\nu_T^{\circ}$. {We see that the optimal
  eddy viscosities $\nu_T^{\bullet}$ obtained both with short and long
  optimization windows allow the Galerkin model to track the target
  {energy} $\widetilde{E}(t)$ more accurately than with the reference
  relation, although in fairness to the latter it has to be recognized
  that the choice of $\nu_T^a$ in \eqref{Eqn:ReferenceEddyViscosity}
  was not optimal resulting in overestimated turbulent kinetic energy.
  In fact, the present approach may be considered a systematic way of
  using data to refine closures proposed based on theoretical or
  empirical arguments. The above observations are confirmed by the
  values of the mean turbulent kinetic energy and its standard
  deviation collected for the different cases in Table \ref{tab:ab}.
  We note, in particular, that with optimization performed over long
  time windows the proposed approach captures the mean energy {of the
    flow} with the accuracy of two significant digits.  The average
  modal energies $\overline{E}_i$, $i=1,\dots,100$, are presented in
  figure \ref{fig:Ktab}b and we see that the optimal eddy viscosity
  $\nu_T^{\bullet}$ yields an improved reconstruction essentially
  across the entire mode spectrum. This should be contrasted with
  figure \ref{fig:Ktml}b, where an improvement was observed only for
  the first energy-containing modes.  This difference is {attributed}
  to the spectral properties of the two flows and our choice of the
  cost functional \eqref{Eqn:J} --- while in the mixing-layer flow
  most of the flow energy is contained in the first few modes, in the
  Ahmed body wake this energy is spread over a very large number of
  modes.  These findings are corroborated by the plots of the
  time-histories of selected Galerkin coefficients $a_i(t)$,
  $i=1,5,25,100$, shown in figure \ref{fig:aiab}.  In those plots we
  note that, unlike the case of the mixing layer, some improvement is
  also obtained for higher modes. Finally, the correlation functions
  \eqref{Eqn:Corr}--\eqref{Eqn:Corr2} for the POD projections of the
  original flow data and the solutions of the reduced-order model
  \eqref{Eqn:GalerkinSystemEddyViscosity} with the reference and
  optimal eddy viscosities are shown in figure \ref{fig:corrab}.  All
  curves reveal a small oscillatory component corresponding to the von
  K\'arm\'an vortex shedding at the Strouhal number $St_H\approx 0.2$.
  These oscillations are not very pronounced in the velocity fields,
  but show up more clearly in the pressure field and the aerodynamic
  forces as reported by \citet{Osth2014jfm}. The curve corresponding
  to the LES data shows an anti-correlation after roughly 100
  convection times.  This behaviour can be traced back to the
  asymmetric base flow drift from a state with positive to a state
  with negative transverse forces.  This base flow drift is resolved
  by the shift mode ($a_1$ in Figure \ref{fig:aiab}). From the same
  plot of the POD mode coefficients, the reduced-order models are seen
  to display a smaller base flow variation than exhibited by the
  actual LES data. This explains the decreased variation of the
  correlation function of the POD models.  We emphasize that it is
  very difficult for POD models to resolve multi-scale phenomena, such
  as vortex shedding combined with base flow drifts, the time-scales
  of which are two orders of magnitude apart. For further details
  concerning the reduced-order modelling of the Ahmed body wake, we
  refer the reader to the original publication by
  \citet{Osth2014jfm}.}

\section{Conclusions and future directions}
\label{Sec:Conclusions}

We have proposed an optimal nonlinear eddy viscosity relation for a
large class of reduced-order models which improves on the results from
a number of earlier studies.  In the pioneering investigation
concerning POD-based reduced-order models by \citet{Aubry1988jfm}, a
single constant eddy viscosity parameter was assumed.
\citet{Rempfer1994jfm2} proposed a mode-dependent refinement of the
constant eddy viscosity ansatz which significantly improves the
accuracy of reduced-order models. Later, \citet{Noack2011book} derived
a nonlinear eddy viscosity as a function of the square root of the
resolved fluctuation energy in which constant ratios between the modal
energies were assumed.  This nonlinearity guarantees the boundedness
of the Galerkin solution \citep{Cordier2013ef}.  As shown by
\citet{Osth2014jfm}, combinations of modal and nonlinear eddy
viscosities may improve the accuracy and robustness of POD-based
reduced-order models. The key new aspect of the approach proposed here
is that the eddy viscosity relations are defined to be optimal in a
mathematically precise sense. {As such, these relations can be
  viewed as systematic, data-based refinements of closures obtained
  from theoretical or empirical considerations.}

The current study addresses the limitations of earlier approaches by
considering the eddy viscosity as an arbitrary function of the
resolved turbulent kinetic energy which is optimized by matching the
fluctuation level of the reduced-order model to the corresponding
quantity of the reference data.  This optimization is performed with a
generalization of the 4D-VAR data assimilation method adopted for the
reconstruction of constitutive equations by \citet{bvp10,bp11a}.

POD models with the optimal eddy viscosity are constructed for three
shear flows with progressively richer dynamics {spanning {the}
laminar and turbulent regimes}.  First, the two-dimensional POD model
for the transient behaviour in the 2D cylinder wake is recalled from
an earlier study \citep{pnm14}.  Here, a negative eddy viscosity is
derived at low fluctuation levels to compensate for the significantly
underpredicted growth rate of the POD model {(figure \ref{fig:g1l}b)}.
On the other hand, on the limit cycle and beyond, a positive eddy
viscosity models the energy transfer to the higher-order modes.  In
this example, the eddy viscosity not only assures correct amplitudes
on the limit cycle, but also yields more accurate transient times
{(figure \ref{fig:g1l}a)}.

Second, a 20-dimensional POD model of the 2D mixing layer at $Re=500$
with velocity ratio $3$ is investigated.  The starting point was a
reduced-order model with a single nonlinear eddy viscosity calibrated
against a DNS of the Navier-Stokes system by \cite{Cordier2013ef}.
Good agreement between the POD model and the DNS was observed with
respect to the {frequency content and the modal fluctuation levels for
{the} energy-containing modes (figures \ref{fig:Ktml}b and
\ref{fig:aiml})}.  Surprisingly, the optimal eddy viscosity
significantly deviates from the square-root ansatz
\eqref{Eqn:ReferenceEddyViscosity} and attains negative values for a
range of low fluctuation levels, thus accelerating the slow transients
of the reduced-order model \citep{Noack2005jfm}. However, the optimal
eddy viscosity $\nu_T^{\bullet}$ is larger than the square-root ansatz
$\nu_T^{\circ}$ at larger fluctuation levels thus limiting more
energetic events {(figure \ref{fig:nuTml})}.

Third, a 100-dimensional POD model of the 3D Ahmed body wake at
Reynolds number $300,000$ is constructed.  The starting point is a
large eddy simulation and the best one from the Galerkin POD models
developed and analyzed by \citet[model ``GS-D'']{Osth2014jfm} is used
as a benchmark. The sub-scale turbulence representation in this model
includes the modal eddy viscosities proportional to the square-root of
the resolved turbulent kinetic energy (cf.~\S\
\ref{Sec:EddyViscosity}).  The optimal eddy viscosity respects the
ratio between the modal viscosities while allowing for an arbitrary
scaling with the resolved turbulent kinetic energy. As regards the
comparison between the optimal and reference eddy viscosity {(figure
\ref{fig:nuTab})}, {for} all values of the fluctuation energy the
optimal eddy viscosity $\nu_T^{\bullet}$ exhibits larger values than
the reference relation $\nu_T^{\circ}$, consistently with the
overprediction of the energy fluctuation level in the latter case
{(figure \ref{fig:Ktab}a)}.

{A key advantage of variational optimization formulations such as the
one proposed here is that they reveal certain performance trade-offs
inherent in the solution of complex flow problems which can hardly be
identified based on the physical intuition alone. This is evident when
one compares the results obtained in the mixing-layer flow, which can
be considered ``laminar'', and the Ahmed body wake, which is
``turbulent'' in all respects. Since in the first case most of the
turbulent kinetic energy was associated with the first two POD modes,
these were also the components of the ROM mostly affected by the
optimization process (figure \ref{fig:Ktml}b). On the other hand, in
the second case, in which the energy was distributed more evenly among
different modes, optimization affected the entire spectrum (cf.~figure
\ref{fig:Ktab}b). This comparison demonstrates that the optimal eddy
viscosities do indeed adapt to situations characterized by
{essentially} different flow physics. {At the same time,
  these results also reveal certain fundamental performance
  limitations inherent in the ansatz $\nu_T = \nu_T(E)$ commonly used
  for the eddy viscosity.}  Needless to say, this process can be
modified by using a different cost functional {and/or a different
  ansatz for $\nu_T$}. For example, adopting a cost functional
penalizing deviations of, say, enstrophy rather than energy, would
have certainly yielded different results.  We emphasize that such
decisions are a part of the problem formulation and can be handled by
the proposed solution approach in a straightforward manner.}

The observed features of the optimal eddy viscosity identified as a
function of the fluctuation energy deserve additional discussion.
From the results we conjecture that the optimal eddy viscosity
$\nu_T^{\bullet}$ does not strongly depend on the chosen time window
$[0,T]$, provided that it is equal to or longer than the
characteristic time scale of the dominant coherent structures.  This
was the case for both of the time windows used for the Ahmed body wake
{(figure \ref{fig:Ktab}a)}, but not for the short time window
used for the mixing layer {(figure \ref{fig:Ktml}a)}.  Secondly,
the eddy viscosity obtained for the mixing layer shows two minima
helping stabilize two different energy levels {(figure
  \ref{fig:nuTml})}.  This bimodal behaviour is consistent with the
cluster-based analysis of the same data performed by
\citet{Kaiser2014jfm}. {It is shown there} that the mixing layer
flow has two quasi-attractors: one which is dominated by the
Kelvin-Helmholtz instability at a lower energy level and another one
dominated by period-doubling at a higher energy level.  Thirdly, the
mixing layer model exhibits a negative eddy viscosity while the model
of the Ahmed body flow does not.  We conjecture that this difference
has two reasons: the first is that the fluctuation levels of the
mixing layer have relatively larger variations {(figure
  \ref{fig:Ktml}a)}, hence we can estimate transient times for this 2D
flow better than for the 3D wake; the second is that a negative eddy
viscosity excites coherent structures with similar scales in the POD
model of the mixing layer flow. On the other hand, for the Ahmed body
wake, a negative eddy viscosity would imply that the strongly damped
high-order modes would suddenly become excited which would in turn
lead to an unphysical inverse energy cascade.  Summarizing, the
different features of the optimal eddy viscosity found for the 2D and
3D shear flows are consistent with our expectations based on the
behaviour of POD models.

Concerning the choice of the parameters in the optimization
formulation, we note that the cost functional tracking the error of
the fluctuation energy gives quite comparable results over different
time windows (cf.~figure \ref{fig:subintervals}), provided that the
windows cover at minimum several characteristic flow periods.  This
was the case for the Ahmed body flow in which the shedding period was
5-10 time units, whereas optimization was performed over intervals
with $\Delta T = 20$ and $\Delta T = 200$ {(figure \ref{fig:Ktab}a)}.
On the other hand, for the mixing layer the shorter window with
$\Delta T = 10$ covered only {about} half of the Kelvin-Helmholtz
shedding period {(figure \ref{fig:Ktml}a)} and the resulting optimal
eddy viscosity was significantly different from the relations found by
solving optimization problem \eqref{Eqn:P} with subintervals 10 times
longer {(figure \ref{fig:nuTml}). As regards Assumption \ref{ass1} and
its validity, we remark that statements {(a) and (b)} are
mathematical in nature and ensure that model
\eqref{Eqn:GalerkinSystemEddyViscosity} is well-posed. Statements
{(c) and (d)} stipulate that for values of $E$ for which the
sensitivity information is not available the optimal eddy viscosity
$\nu_T^{\bullet}(E)$ should revert to some chosen reference relation,
in our case relation \eqref{Eqn:ReferenceEddyViscosity}.}

In providing a closure relation for unresolved fluctuations based on
solution data, the proposed approach to identifying the optimal eddy
viscosity bears some resemblance to the ``optimal LES'' methodology
which originated with \citet{lm99}. However, it differs from the
optimal LES in that our optimal eddy viscosity $\nu_T^{\bullet}$ is
reconstructed in a non-parametric manner.  The proposed closure
strategy can be employed in a straightforward manner to identify
closure relations depending on one variable for a large class of
reduced-order models {both for laminar and turbulent flows}. A
highly relevant problem complementary to the problem solved in this
study is optimization of the dependence of the eddy viscosity on the
mode index $i$ while keeping the dependence on the turbulent kinetic
energy fixed. The approach developed here can be adapted to solve such
problems by treating the discrete mode index $i$ as a continuous
variable, i.e., an effective wavenumber of the mode.  {It will be
  interesting to see whether such a formulation can lead to improved
  performance with respect to the ansatz $\nu_T = \nu_T(E)$ used in
  the present investigation.}  This problem will be studied in the
near future.  Another related problem concerns determination of
optimal turbulence closure strategies for simplified flow models
defined in the PDE setting such as the RANS and LES approaches (in
fact, these are the type of problems the reconstruction method we used
was initially developed for, see \citet{bvp10,bp11a}).  As regards LES
models, an interesting open problem is determination of optimal wall
damping functions \citep{d56}.  Problems of such type also arise in
fundamental turbulence research, for example, in the context of the
{K\'arm\'an}-Howarth equation.  Other, possibly less obvious,
extensions of this methodology include optimal identification of
inertial manifolds and feedback control laws, and the authors are
already pursuing these applications in the context of closed-loop
turbulence control.

\section*{Acknowledgements}

The authors thank Shervin Bagheri, Laurent Cordier and {Sini\v{s}a
  Krajnovi\'c} for stimulating discussions and for providing the data
used in figure \ref{fig:Ktml2} (L.C.).  Funding for this research was
provided by the French Agence Nationale de la Recherche (ANR) via the
Chair of Excellence TUCOROM and is gratefully acknowledged.  B.P. was
also partially supported by a Discovery Grant from the Natural
Sciences and Engineering Research Council of Canada (NSERC). A part of
this work was based on J.~\"Osth's Ph.D.~thesis which was financially
supported by Trafikverket (Swedish Transport Administration).  {The
authors also thank the referees for their detailed, thoughtful and
helpful suggestions.}


\appendix
\section{Derivation of Gradient Expression}
\label{Sec:Adjoint}

In this appendix we derive expression \eqref{Eqn:gradL2} for the $L_2$
gradient of cost functional \eqref{Eqn:J}. The key observation is
that, since the G\^{a}teaux differential of $\J(\nu_T)$ appearing in
\eqref{Eqn:opt} is a bounded linear functional with respect to its
second argument $\nu_T' \in \X(\I)$, $\X(\I)$ being an appropriate
Hilbert function space, by the Riesz representation theorem
\citep{b77} we have
\begin{equation}
\forall_{\nu_T' \in \X(\I)} \quad 
\J'(\nu_T;\nu_T') = \Big\langle \nabla^{\X}\J, \nu_T' \Big\rangle_{\X(\I)},
\label{Eqn:Riesz}
\end{equation}
where $\langle \cdot, \cdot \rangle_{\X(\I)}$ denotes the inner
product in the space $\X(\I)$. We identify the Riesz representer
$\nabla^{\X}\J$ as the {\em gradient} of $\J$ with respect to the
topology of the space $\X(\I)$ (in the present problem, we have either
$\X(\I) = L_2(\I)$ or $\X(\I) = H^1(\I)$). We begin by computing the
G\^{a}teaux differential of \eqref{Eqn:J} which yields
\begin{equation}
\J'(\nu_T;\nu_T') = \frac{1}{T} \int_0^T \left[ E(t) - \widetilde{E}(t) \right] E'(t) \, dt 
=  \frac{1}{T} \int_0^T \left[ E(t) - \widetilde{E}(t) \right] \sum_{i=1}^N a_i(t) a'_i(t) \, dt,
\label{Eqn:dJ}
\end{equation}
where $E' := \sum_{i=1}^N a_i a'_i$ and $a'_i(t)$, $i=1,\dots,N$,
solve the linearization of system \eqref{Eqn:GalerkinSystem4}.
Following the approach described by \citet{pnm14}, this linearization
can be shown to have the form
\begin{subequations}
\label{Eqn:LinGalerkinSystem}
\begin{align}
& \frac{da'_i}{dt} = \sum_{j=0}^N \left[ \sum_{k=0}^N (q_{ijk}+q_{ikj})a_k  + l^{\nu}_{ij} \left( \nu + \nu_T(E(t)) + \frac{d \nu_T}{de} a_ia_j\right)\right] a'_j 
+ \nu'_T(E(t)) \, \sum_{j=0}^N l^{\nu}_{ij} a_j, \nonumber \\
& \phantom{\frac{da'_i}{dt}} =: \sum_{j=0}^N A_{ij} a'_j + \nu'_T(E(t)) \, \sum_{j=0}^N l^{\nu}_{ij} a_j,
\qquad t \in ((m-1)\Delta T, m\Delta T], \label{Eqn:LinGalerkinSystema} \\
& a'_0(t) = 0,  \label{Eqn:LinGalerkinSystemb} \\
& a'_i((m-1)\Delta T) = 0, \qquad i=1,\dots,N,\quad m=1,\dots,M,
\label{Eqn:LinGalerkinSystemc}
 \end{align}
\end{subequations}
where the second line in \eqref{Eqn:LinGalerkinSystema} defines the
linear operator $\bA$. We note that differential \eqref{Eqn:dJ} is not
yet in a form consistent with Riesz representation \eqref{Eqn:Riesz},
because the perturbation variable $\nu'_T$ does not enter as a linear
factor in \eqref{Eqn:dJ}, but is instead hidden in a source term in
equation \eqref{Eqn:LinGalerkinSystema}. In order to transform
\eqref{Eqn:dJ} into Riesz form \eqref{Eqn:Riesz}, we introduce the
{\em adjoint state} $\a^*(t) = [0,a^*_1(t),\dots,a^*_N(t)]^T \in
\RR^{N+1}$, so that integrating it against the perturbation equation
\eqref{Eqn:LinGalerkinSystema} and applying integration by parts we
obtain
\begin{equation}
\begin{aligned}
& \sum_{i=1}^N \, \int_0^T \left(\frac{da'_i}{dt} -  \sum_{j=0}^N A_{ij}a'_j - \nu'_T(E(t)) \, \sum_{j=0}^N l^{\nu}_{ij} a_j \right)a^*_i \, dt = \\
 \sum_{i=1}^N a'_i a^*_i \big|_{t=0}^{t=T} +
& \sum_{i=1}^N \, \int_0^T a'_i \left(-\frac{da^*_i}{dt} -  \sum_{j=0}^N A_{ji}a^*_j \right) \, dt 
- \int_0^T \nu'_T(E(t)) \, \sum_{i=1,j=0}^N l^{\nu}_{ij} a_j a^*_i \, dt = 0.
\end{aligned}
\label{Eqn:I1}
\end{equation}
Since $a_0^*(t) \equiv 0$, summation over index $i$ in \eqref{Eqn:I1}
starts at 1.  Defining the {\em adjoint system} as in
\eqref{Eqn:GalerkinAdjoint1}, and using it together with
\eqref{Eqn:dJ} and \eqref{Eqn:LinGalerkinSystem}, we obtain from
\eqref{Eqn:I1}
\begin{equation}
\J'(\nu_T;\nu_T') = \int_0^T \nu'_T(E(t)) \, \sum_{i,j=0}^N l^{\nu}_{ij} a_j a^*_i \, dt.
\label{Eqn:dJ2}
\end{equation}
In order to transform this expression to the Riesz form induced by
$\X(I) = L_2(\I)$, i.e.,
\begin{equation}
\J'(\nu_T;\nu_T') = \int_0^{\Em} \nabla^{L_2}\J \, \nu_T' \, de,
\label{Eqn:RieszL2}
\end{equation}
we need to change the integration variable in \eqref{Eqn:dJ2} from
time $t$ to turbulent kinetic energy $e$
\begin{equation}
\begin{aligned}
& \frac{de}{dt} = \sum_{i=1}^N a_i \frac{da_i}{dt} 
= \sum_{i=1}^N a_i \left( f_i(\a) + \nu_T(E(t)) \, \sum_{j=0}^N l^{\nu}_{ij} a_j \right) \\
& \Longrightarrow \quad dt = \frac{de}{\sum_{i=1}^N a_i \left[ f_i(\a) + \nu_T(E(t)) \, \sum_{j=0}^N l^{\nu}_{ij} a_j \right]},
\end{aligned}
\label{Eqn:t2e}
\end{equation}
so that G\^ateaux differential \eqref{Eqn:dJ2} becomes
\begin{align}
\J'(\nu_T;\nu_T') 
& = \int_{\C} \frac{\sum_{i,j=0}^N l^{\nu}_{ij} a_j a^*_i}{\sum_{i=1}^N a_i \left[ f_i(\a) + \nu_T(E(t)) \, \sum_{j=0}^N l^{\nu}_{ij} a_j \right]} \, \nu'_T(e) \, de \nonumber \\
& = \int_0^{\Em}  \sum_{\stackrel{t}{E(\a(t)) = e}} \frac{\sum_{i,j=0}^N l^{\nu}_{ij} a_j a^*_i}{\sum_{i=1}^N \big| a_i  f_i(\a) + \nu_T(E(t)) \, \sum_{j=0}^N l^{\nu}_{ij} a_i a_j \big|} \, \nu'_T(e) \, de,
\label{Eqn:dJ3}
\end{align}
where the first expression on the right-hand side in \eqref{Eqn:dJ3}
is an integral over the system trajectory $\C$ in the phase space
$\RR^n$ (i.e., a {\em line} integral in which $de$ can be either
positive or negative), whereas the second expression is a {\em
  definite} integral consistent with Riesz form \eqref{Eqn:RieszL2}.
Thus, identifying \eqref{Eqn:dJ3} with \eqref{Eqn:RieszL2}, we finally
obtain gradient expression \eqref{Eqn:gradL2}.



\end{document}